\documentclass[prb,aps,showpacs,twocolumn,unsortedaddress]{revtex4}
\usepackage{graphics, bm}
\usepackage{psfrag}
\usepackage{amsmath}
\usepackage{amssymb}
\usepackage{epsfig}
\usepackage{subfigure}
\usepackage{grffile}
\usepackage{color}
\usepackage{morefloats}
\newcommand{\beq}    {\begin{equation}}
\newcommand{\enq}    {\end{equation}}
 \newcommand{\ceq}[1] {(\ref{#1})}
\newcommand{\kk}{{\bf k}}
\newcommand{\rr}{{\bf r}}
\newcommand{\qq}{{\bf q}}
\newcommand{\sio}    {${\rm SiO_2}\;$}
\DeclareMathOperator{\sgn}{sgn}
\newcommand{\df}     {\equiv}
\newcommand{\nimp}   {\rm{n}_{\rm imp}}
\newcommand{\nrms}   {\rm{n}_{(\rm rms)}}
\newcommand{\nrmsone}   {\rm{n}_{1(\rm rms)}}
\newcommand{\nrmstwo}   {\rm{n}_{2(\rm rms)}}
\newcommand{\vrms}   {\rm{V}_{{\rm sc} \; (\rm rms)}}
\newcommand{\vrmsa}   {\rm{V}^{(1)}_{{\rm sc}(\rm rms)}}
\newcommand{\vrmsb}   {\rm{V}^{(2)}_{{\rm sc}(\rm rms)}}
\newcommand{\nrmsa}   {\rm{n}_{1(\rm rms)}}
\newcommand{\nrmsb}   {\rm{n}_{2(\rm rms)}}
\newcommand{\nav}     {\langle \rm{n} \rangle}
\newcommand{\nava}    {\langle \rm{n}_{1} \rangle}
\newcommand{\navb}    {\langle \rm{n}_{2} \rangle}
\newcommand{\aia}     {A_I^{(1)}}
\newcommand{\aib}     {A_I^{(2)}}

\newcommand{\Deltaav}    {\langle \Delta \rangle}

\newcommand{\Deltarms}   {\langle \Delta_{\rm rms} \rangle}

 \newcommand{\rb}     {r_{\rm sc}}

\graphicspath{{../../Figures/}}

\begin{document}

\title{Ground state of graphene heterostructures in the presence of random charged impurities.}

\author{Martin Rodriguez-Vega$^1$, Jonathan Fischer$^1$\footnote{Present address: Department of Statistics, University of California at Berkeley, Berkeley, CA 94720-3860, USA.}, S. Das Sarma$^2$, and E. Rossi$^1$}
\affiliation{
             $^1$Department of Physics, College of William and Mary, Williamsburg, VA 23187, USA\\
             $^2$Condensed Matter Theory Center, Department of Physics, University of Maryland, College Park, Maryland 20742-4111, USA
            }
\date{\today}

   
\begin{abstract}
We study the effect of long-range disorder created by charge impurities on the carrier density distribution of graphene-based heterostructures.
We consider heterostructures formed by two graphenic sheets (either single layer graphene, SLG, or bilayer graphene, BLG)
separated by a dielectric film. We present results for symmetric heterostructures, SLG-SLG and BLG-BLG, and 
hybrid ones, BLG-SLG. 
As for isolated layers, we find that the presence of charged
impurities induces strong carrier density inhomogeneities, especially
at low dopings where the density landscape breaks up in electron-hole
puddles. 
We provide quantitative results for the strength of the carrier density inhomogeneities
and for the screened disorder potential for a large range of experimentally relevant conditions.
For heterostructures in which BLG is present we also present results 
for the band-gap induced by the 
perpendicular electric field generated self-consistently by the disorder potential and by the distribution
of charges in the heterostructure.
For SLG-SLG heterostructures we discuss the relevance of our results for the understanding of
the recently observed metal-insulator transition in each of the graphene layers
forming the heterostructure.
Moreover, we calculate the correlation between the density profiles in the two graphenic layers
and show that for standard experimental conditions the two profiles
are well correlated.
\end{abstract}


\maketitle

\section{Introduction}
The ability to realize single layer graphene (SLG) \cite{novoselov2004}, bilayer graphene (BLG)
\cite{novoselov2006}, 
and other two-dimensional (2D) crystals \cite{novoselov2005b},
combined with recent advances in fabrication techniques
\cite{li2009,ismach2012} in recent years
has allowed the realization of novel 2D heterostructures 
\cite{kim2009,kim2012,haigh2012,hunt2013,britnell2012,britnell2013,lee2013,gamucci2014,dang2010,song2010,jin2013,jzhang2013b}.
In these structures, two or more 2D crystals are stacked in a designed sequence.
Layers of hexagonal boron nitride (hBN)
\cite{dean2010,xue2011,yankowitz2012}
have been used to electrically separate 
the graphenic layers (SLG or BLG) in multilayered 2D heterostructures.
In particular, hBN allows the realization of graphene-based heterostructures
in which the graphenic layers are very close and yet electrically separated
\cite{gorbachev2012,geim2013}, a situation that is ideal to study the effects of interlayer
interactions. 
It has been proposed that in these type of systems the interlayer interactions
can drive the system into spontaneously broken symmetry ground states
\cite{min2008b, zhang2008jog, kharitonov2008b, kharitonov2010, zhang2010, jzhang2013}.
So far, experiments have 
not observed clear signatures of the establishment of these
collective ground states. However, recent measurements of
the drag resistivity in graphene double layers
\cite{gorbachev2012}
have shown that the drag resistivity has a very large and anomalous
peak when the doping in both graphene sheets is set to zero.
This phenomenon indicates that a strong correlation is present between
the carriers in the two layers. 

%
In most of the samples random charge impurities are present in the graphene
environment, either in the substrate or trapped between the graphenic
layer and the substrate. It has been shown theoretically 
\cite{rossi2008} and experimentally 
\cite{martin2008, zhang2009, deshpande2009,deshpande2009b}
that the long-range disorder due to charge impurities induces strong, long-range, carrier density inhomogeneities in isolated SLG and BLG.  
The presence of random carrier density inhomogeneities has been predicted theoretically to strongly suppress the critical temperature ($T_c$)
for the formation of an interlayer phase coherent state
\cite{abergel2012c,abergel2013, Efimkin2011} in graphene heterostructures. This is in contrast to the short-range disorder that 
is not expected to suppress significantly $T_c$
\cite{bistritzer2008a,jzhang2013,jzhang2014}.
In addition, the presence of charge inhomogeneities, correlated
in the two layers, is a necessary ingredient of the energy-transfer mechanism that has been proposed
\cite{song2012,song2013}
to explain the strong peak of the drag-resistivity at the double-neutrality point.
Disorder-induced carrier density inhomogeneities
are also expected to strongly affect the transport properties of graphene-based
heterostructures 
\cite{hwang2007,adam2007,rossi2009,fogler2009,dassarma2011,rossi2012}.
For these reasons, the accurate characterization of the carrier density inhomogeneities
induced by long-range disorder in graphene-based heterostructures is essential
to understand the fundamental properties of these systems and to identify ways
to increase their electronic mobility.

The characterization of the effects of disorder in graphene-based heterostructures
is challenging for several reasons:
(i)   In most samples the disorder appears to be due predominantly to random charge impurities and to
      be quite strong and long-range, this fact
      makes the use of standard techniques, such as perturbation theory, not viable;
(ii)  Due to the linear dispersion in graphene, the screening of the long-range disorder due to the charge impurities
      is nonlinear;
(iii) In graphene heterostructures the screening effects due to the different layers must be 
      taken into account self-consistently;
(iv)  In bilayer graphene the presence of a perpendicular electric field opens a band-gap \cite{oostinga2007,zhang2009c}; 
(v)   In heterostructures comprising BLG the component of the electric field perpendicular to BLG, and the BLG gap, must be obtained
      self-consistently taking into account the presence of the disorder and its screening by the metallic gates, and the other graphenic layer.
In this work we present a systematic study of the effects of the long-range disorder
due to random charge impurities on the ground state of graphene-based heterostructures taking into account all the effects mentioned above.
As shown in Fig.~\ref{fig:setup} we consider heterostructures formed by two ``graphenic'' layers, either SLG or BLG,
separated by a thin dielectric film. In the assumed configuration, using a top and a bottom gate, the 
doping of each graphenic layer can be set independently. We considered three classes of heterostructures:
(i)   double layer graphene (SLG-SLG) formed by two sheets of single layer graphene;
(ii)  double bilayer graphene (BLG-BLG) formed by two sheets of bilayer graphene;
(iii) ``hybrid structures'' (BLG-SLG) formed by one sheet of BLG and one sheet of SLG.
We find that the presence of charge impurities induces strong and long-range carrier density
inhomogeneities in graphene-based heterostructures as 
in isolated SLG \cite{rossi2008} and BLG \cite{rossi2009}. However, for typical experimental situations
we find that for the top graphenic layer the strength of the carrier density inhomogeneities
is strongly suppressed due to the screening of the charge impurities by the bottom layer.
We quantify this effect for most of the experimentally relevant conditions and find that
for the top layer the amplitude of the density fluctuations can be reduced by an order of magnitude and that the effect is strongest in BLG-SLG heterostructures.
We also show that the carrier density inhomogeneities in the different graphenic
layers are well correlated. Finally, we show how the average band gap of BLG and its root mean square
depend on the parameters, such as the impurity density, characterizing the heterostructure.
Our results present a comprehensive characterization of the carrier density profile
of graphene heterostructures in the presence of long-range disorder. By showing how
the strength of the carrier density inhomogeneities depend on the experimental parameters,
our results show how the quality of
graphene-based heterostructures could be improved.
In particular,
the parameters that, within a certain range, can be easily tuned experimentally are:
the doping of each of the graphenic sheets forming the heterostructure,
the type of graphenic sheets used, the impurity density (via annealing
or the use of different substrates), and the distance
between the graphenic sheets. For each of these parameters we present quantitative results that show how the values
can be tuned to reduce the disorder strength in each of the  graphenic sheets, or both,
forming the heterostructure.
The results presented in the remainder of this work,
for instance, quantify how an increase of the doping in one of the two sheets forming
the heterostructure can substantially reduce the strength of the disorder-induced long-range inhomogeneities
in the other sheet, and  quantify how much a reduction of the impurity density would reduce
the strength of the disorder potential in the heterostructure. 
In addition, we show how a change of the distance between the two sheets
can be optimized to reduce the overall disorder strength in the heterostructure. 
Our results also show that, to reduce the 
strength of the disorder-induced long-range inhomogeneities in
single layer graphene it is more efficient to have below it a sheet of bilayer
graphene instead of SLG. 
The information on how to reduce, control, the disorder strength is essential for the study of fundamental effects in 
graphene heterostructures and for their use in technological applications.

In section \ref{sec:method} we present our theoretical approach;
in section \ref{sec:results} we present the results and discuss their relevance
for current experiments; 
in section \ref{sec:mit} we discuss the relevance of our results for
the recently observed metal-insulator transition as a function of doping in 
double-layer graphene heterostructures.
Finally in section \ref{sec:conclusions} we present our conclusions.

\section{Theoretical approach}
\label{sec:method}
\begin{figure}[!th]
 \begin{center}
  \centering
  \includegraphics[width=6.5cm]{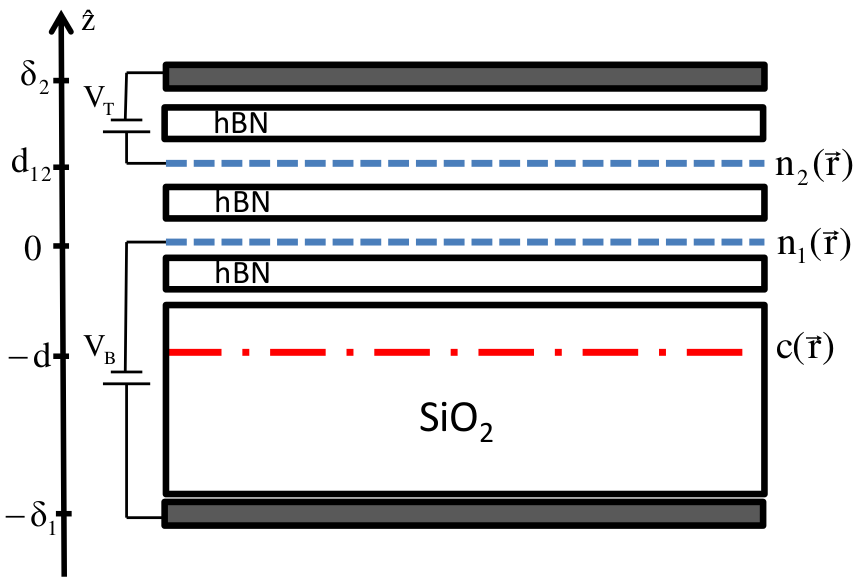}
  \caption{
    Sketch of the typical graphene heterostructure considered in this work showing the graphenic layers (blue dashed lines) connected to independent metal gates (gray solid lines), isolated with hBN, and placed on a SiO$_2$ substrate. The charged impurities are modeled as a two-dimensional distribution c$(\rr)$ (red line) located at an effective distance $d$ below the bottom graphenic layer.
          } 
  \label{fig:setup}
 \end{center}
\end{figure}

%
Figure~\ref{fig:setup} presents a sketch of the type of graphene heterostructure that we consider.
One graphenic layer (SLG or BLG), layer 1 in our notation is placed on an insulating substrate, typically \sio. 
A thin buffer layer of high quality dielectric, typically hBN, might be present between the \sio and the graphenic layer.
A second graphenic layer, layer 2, is placed above the first one.
Layer 2 and layer 1 are electrically isolated via a thin insulating film.
The doping level of the two graphenic layers can be tuned independently via
a top and a bottom gate. 

There is compelling evidence 
\cite{dassarma2011}
that in systems of the type depicted in Fig.~\ref{fig:setup} 
the dominant sources of disorder are random charge impurities
located close to the surface of \sio. 
%
It is known that on the surface of \sio
there is a large density of charge impurities.
Transport measurements on single layer graphene have consistently observed
a linear scaling of the conductivity with the doping ($n$), at low doping.
The fact that the conductivity is suppressed at low dopings indicates that
the effective strength of the disorder increases as the carrier density 
is decreased. 
Theoretical transport results in which 
charge impurities are the dominant source of scattering precisely predict
at low dopings a linear suppression of the conductivity as $n$ is decreased
\cite{hwang2007,nomura2006,dassarma2011}.
The agreement between transport theories in which charge impurities are
the main source of disorder and experimental transport measurements has also
been confirmed by experiments in which the density of charge impurities
was tuned \cite{chen2008}. 
In recent years there have been also
several imaging experiments 
\cite{martin2008,zhang2009,deshpande2009,deshpande2009b}
that, close to the charge neutrality point,
have observed the presence of electron-hole puddles with dimensions and amplitudes
that are consistent with the presence of charge impurity densities in the 
graphene environment \cite{adam2007,hwang2007,rossi2008} of the order
of the ones extracted from the transport results mentioned above.

The distribution of the charge impurities can be modeled as  an effective 2D distribution $c(\rr)$ placed
at a distance $d$ below the bottom graphenic layer (layer 1).
The dash-dot line in Fig.~\ref{fig:setup} shows
schematically the location of the effective 2D plane where the random impurities
are located. It is likely that some charge impurities will also be trapped between
each graphenic sheet and the adjacent thin dielectric films. However,
experimental evidence, especially for setups in which hBN is used as
dielectric material, strongly suggests that the density of such trapped 
impurities is at least an order of magnitude smaller that the density
of the impurities close to the surface of the \sio. For this reason
we henceforth assume that the disorder potential is solely 
due to the charge impurities located close to the ${\rm SiO}_2$'s surface.
Without loss of generality, we can assume $\langle c(\rr)\rangle=0$,
where the angle brackets denote average over disorder realizations.
Our formalism allows to easily take into account the presence
of spatial correlation between the charge impurities \cite{li2011,li2012}.
However, given the fact that in general the charge impurities are frozen and locked
in a configuration that results from the fabrication process and that is not
the thermodynamic equilibrium \cite{yan2011}, we can assume 
that their position is uncorrelated 
so that $\langle c(\rr) c(\rr')\rangle = \nimp\delta(\rr-\rr')$, where $\nimp$
is the charge impurity density.

At low energies the fermionic excitations of SLG are well described 
by a  massless Dirac model with Hamiltonian \cite{neto2009,dassarma2011}:
\beq
H = \hbar v_F \; \bf{ \sigma} \cdot \kk \; ,
\enq
where $\hbar\kk$ is the momentum operator, 
$\bf {\sigma} = (\sigma_x, \sigma_y)$ are the Pauli matrices in sublattice space, 
and  $ v_F\approx 10^6 $ ms$^{-1}$ is the Fermi velocity. 
%
%
Recent experiments for graphene on hBN have shown evidence of the opening of a gap
\cite{hunt2013,amet2013}.
Considering that the fact that there is a 1.8\% lattice mismatch between 
graphene and hBN and the fact that in current experiments a twist angle
between the graphene layer and the hBN is normally present, the mechanism
by which the gaps open is still not completely understood 
\cite{song2013,jung2014}, but is thought to be arising from the explicit breaking of the 'AB' sub-lattice symmetry in SLG due to the presence of the hBN substrate,
and that it should not depend on the local electric field, but should depend on the twist angle between graphene and hBN in some complex manner.
For our purposes this means that for SLG on hBN the band-gap, if present,
can be assumed to be fixed and independent of the local doping and electric
field created by the nearby gates. In the presence of a band gap
the low-energy Hamiltonian for single layer graphene becomes:
\beq
H = \begin{pmatrix} 
      \Delta     & \hbar v_F (k_x-ik_y) \\ 
      \hbar v_F (k_x+ik_y) & -\Delta 
    \end{pmatrix}.
 \label{eqn:gappedslg}
\enq

At low energies the effective Hamiltonian describing the fermionic excitations in BLG is
\beq
H = \begin{pmatrix} 
      \Delta     & \frac{\hbar^2}{2m^*}(k_x-ik_y)^2 \\ 
      \frac{\hbar^2}{2m^*}(k_x+ik_y)^2 & -\Delta 
    \end{pmatrix} \; ,
 \label{eq:Hblg}
\enq
where m$^*=0.033$m$_e$ is the effective electron mass and
$\Delta$ is the band gap due a difference ($U$) in the electrochemical
potential between the two layers of carbon atoms forming BLG.

In our case $U$ in Eq.~\ceq{eq:Hblg}
is due to the presence of a perpendicular electric
field $E_\perp$ induced by the metal gates, the other
graphenic layer, and the charge impurities surrounding
the BLG sheet.
If BLG is layer 1, i.e. it is the graphenic layer closest to 
the charge impurities, we have:
\begin{align}
 E^{(1)}_{\perp}(\rr) & = \frac{e \; d}{\epsilon} \int d\rr' \frac{c(\rr')}{[|\rr-\rr'|^2+d^2]^{3/2}} \nonumber \\
               & - \frac{e \; d_{12}}{\epsilon} \int d\rr' \frac{n_2(\rr')}{[|\rr-\rr'|^2+d_{12}^2]^{3/2}} \nonumber \\
               & - \frac{e\;\delta_1 }{\epsilon} \int d\rr' \frac{n_1(\rr')}{[|\rr-\rr'|^2+\delta_1^2]^{3/2}} \; ,
\label{eqn:perp_efield}
\end{align}
where $d_{12}$ is the distance between the two graphenic layers and 
$\delta_1 \approx 300$nm is the distance between BLG and the bottom gate, Fig.~\ref{fig:setup}.
Notice that in general $E_\perp$ is not uniform, mostly due to the presence of the charge impurities.
When BLG is layer 2 we have:
\begin{align}
E^{(2)}_{\perp}(\rr) & = (d+d_{12})\frac{e}{\epsilon} \int d\rr' \frac{c(\rr')}{[|\rr-\rr'|^2+(d+d_{12})^2]^{3/2}} \nonumber \\
               & + \frac{e \; d_{12}}{\epsilon} \int d\rr' \frac{n_1(\rr')}{[|\rr-\rr'|^2+d_{12}^2]^{3/2}} \nonumber \\
               & + (\delta_2-d_{12}) \frac{e}{\epsilon} \int d\rr' \frac{n_2(\rr')}{[|\rr-\rr'|^2+(\delta_2-d_{12})^2]^{3/2}} \; ,
\label{eqn:twoperp_efield}
\end{align}
where $\delta_2 \approx 150$nm is the distance between the first graphenic layer and the top metal gate, Fig.~\ref{fig:setup}.
Using these expressions for the perpendicular component of the electric field we can calculate $U$.
We have
\beq
U^{(i)}( \rr ) = e d_m E^{(i)}_{\perp}( \rr ) \;,
\label{eqn:potentialenergy}
\enq
where $i=1$ ($i=2$) if BLG is the bottom (top) graphenic layer, and $d_m = 0.335$nm is the BLG interlayer separation.
Taking into account screening effects \cite{min2007,triola2012,abergel2012}
the band gap of BLG due to a finite value of $U$ is given by the equation
\beq
\Delta (x,y) = \frac{\gamma_1 |U(x,y)|}{\sqrt{|U(x,y)|^2+\gamma_1^2}} \; ,
\label{eqn:bandgap}
\enq
where $\gamma_1 = 0.34$~eV is the BLG interlayer tunneling amplitude \cite{neto2009}.

To obtain the ground state carrier density distribution in the presence
of charge impurities we use the Thomas Fermi Dirac theory (TFDT).
The TFDT is a generalization of the Thomas-Fermi theory to include 
cases in which the electronic degrees of freedom behave as massless Dirac
fermions, as in single layer graphene. In this case both the 
kinetic energy functional and the functional due to the exchange
part of the Coulomb interaction are different from those valid
for systems in which the electrons behave as massive fermions
\cite{rossi2008,polini2008}.
In the TFDT the ground state of the system is obtained
by minimizing the energy functional, $E[n]$, of the carrier density $n$.
The TFDT is similar in spirit to the density functional theory (DFT),
the difference being that in the TFDT the kinetic energy is also approximated
by a functional of the density, $E_K[n]$, whereas in the DFT it is treated via the full
quantum-mechanical operator acting on the wave function $\Psi$.
The TFDT returns accurate results as long as the length-scale of
the carrier density inhomogeneities $L_n\equiv|\nabla n/n|^{-1}$ is larger than the Fermi wavelength $\lambda_F$. 
Prior results on SLG \cite{rossi2008,rossi2009} and 
BLG \cite{dassarma2010,rossi2011} have shown that in graphene-based
systems this inequality is satisfied for typical experimental conditions.
The value of $n$ that enters in the
inequality  $L_n\gg\lambda_F$
is the typical local value inside the
``puddles'' characterizing the inhomogenous carrier density landscape.
At the charge neutrality point (CNP) $\nav=0$, however, everywhere
the local density $n(\rr)$ is different from zero and therefore
locally $\lambda_F$ has a finite value.
As a consequence, close to the CNP the 
average density
cannot be taken as a measure of the typical carrier 
density inside the puddles and a better estimate is given by 
the density root mean square $\nrms$.
Given that $\nrms\approx\nimp$ \cite{rossi2008,dassarma2010} 
we have that the TFDT is valid at
all densities as long as $\nimp$ is not too small 
($\nimp > 10^{11} {\rm cm}^{-2}$) \cite{brey2009}.
This is confirmed by prior results on SLG \cite{rossi2008,rossi2009} and 
BLG \cite{dassarma2010,rossi2011}. 
%
The two major advantages of the TFDT are:
%
(i) Being a functional theory is not perturbative with respect to the strength of the density fluctuations and can
       therefore take into account nonlinear screening effects;
(ii) It is computationally very efficient and this makes the TFDT able to return
       disorder-averaged results.
%

\noindent
For the systems of interest,
the TFDT energy functional $E[n_i]$ will
be a functional of
the density profiles, $\{n_i(\rr)\}$, 
in the two graphenic layers.
Neglecting exchange-correlation terms that have been shown to be small
for most of the situation we are interested in \cite{rossi2008,dassarma2010},
the general form of the functional $E[n_i]$ is:
\begin{align}
 E[n_i] = &\sum_{i}E_K[n_i]  +
               \sum_i\frac{e^2}{2 \epsilon} \int d^2 r \int d^2 r' \frac{n_i(\rr)n_i(\rr')}{|\rr-\rr'|} +
               \nonumber \\
              &+\sum_{i,j\neq i}\frac{e^2}{2 \epsilon} \int d^2 r \int d^2 r' \frac{n_i(\rr')n_j(\rr)}{\left[|\rr-\rr'|^2+d_{ij}^2 \right]^{1/2}} 
              \nonumber \\
              &+e\sum_{i}\int d^2 r V_{D}^{i}(\rr)n_i(\rr)  
              -\sum_i\mu_i \int d^2r n_i(\rr)
 \label{eq:En1n2}
\end{align}
where $\epsilon$ is the dielectric constant of the medium surrounding the graphenic layers,
$d_{ij}$ is the distance between the graphenic layers, $V_{D}^{i}$ is the bare disorder potential in layer $i$,
and $\mu_i$ is the chemical potential in layer $i$.
The second term in Eq.~\ceq{eq:En1n2}  is the Hartree part of the intralayer Coulomb interaction, the third
term is the Hartree part of the interlayer Coulomb interaction, and
the fourth is the one due to the disorder potential $V_{D}^{i}$.
Assuming that charge impurities close to the surface of \sio are the dominant source
of disorder we have
\begin{align} 
 &V_D^{(1)} = \frac{e}{\epsilon}\int d\rr'\frac{c(\rr')}{[|\rr-\rr'|^2 + d^2]^{1/2}}; 
 \label{eq:vd1} \\
 &V_D^{(2)} = \frac{e}{\epsilon}\int d\rr'\frac{c(\rr')}{[|\rr-\rr'|^2 + (d+d_{12})^2]^{1/2}}.
 \label{eq:vd2}
\end{align}
%

%
%

The ground state is obtained by minimizing $E$ with respect to $\{n_i\}$.
This gives rise to two coupled equations.
In general, for the cases we are interested in, the term $\mu_{\rm kin}\df\delta E_K/\delta n_i$ 
is nonlinear.
For the case of gapless SLG $\mu_{\rm kin}$ scales as the square-root of the density:
\beq
 \mu^{(SLG)}_{\rm kin}[n] = \hbar v_f \sgn( n(\rr) ) \sqrt{\pi \left| n(\rr) \right|}.
\enq
For the case of gapped SLG we have
\beq
 \mu^{(SLG)}_{\rm kin}[n,\Delta] = \sgn( n(\rr) ) \sqrt{\hbar^2 v^2_f \pi \left| n(\rr) \right| + \Delta^2}
\enq

For BLG, neglecting the presence of a nonzero band-gap ($\Delta$), $\mu_{\rm kin}$ depends linearly on $n$.
This fact allows us to obtain analytical results for the carrier density ground state of BLG-BLG heterostructures
in the limit $\Delta=0$ (see Sec.~\ref{sec:results}).
In the presence of a band gap the screening is strongly non-linear and this is reflected
by the nonlinear dependence of $\mu_{\rm kin}$ with respect to the density.
Taking into account the band-gap for BLG we have
\beq
 \mu^{(BLG)}_{\rm kin}[n] = \sqrt{ \left(\frac{\hbar^2}{2m^*} \right)^2 \pi^2 n^2 + \Delta^2}.
\enq
%
The nonlinearities due to the term $\delta E_K/\delta n_i$, and the need to self-consistently calculate $\Delta$ for systems involving BLG imply that
the solution of the TFDT equations can only be achieved
numerically. We then solve these equations for many (500-1000) disorder
realizations to obtain disorder-averaged results. 
The need to consider many disorder realization to accurately obtain
the disorder-averaged values of the quantities characterizing the ground state
makes the computational efficiency of the TFDT
approach very valuable.

\section{Results}
\label{sec:results}

\begin{figure}[htb]
 \begin{center}
  \centering
   \includegraphics[width=8.5cm]{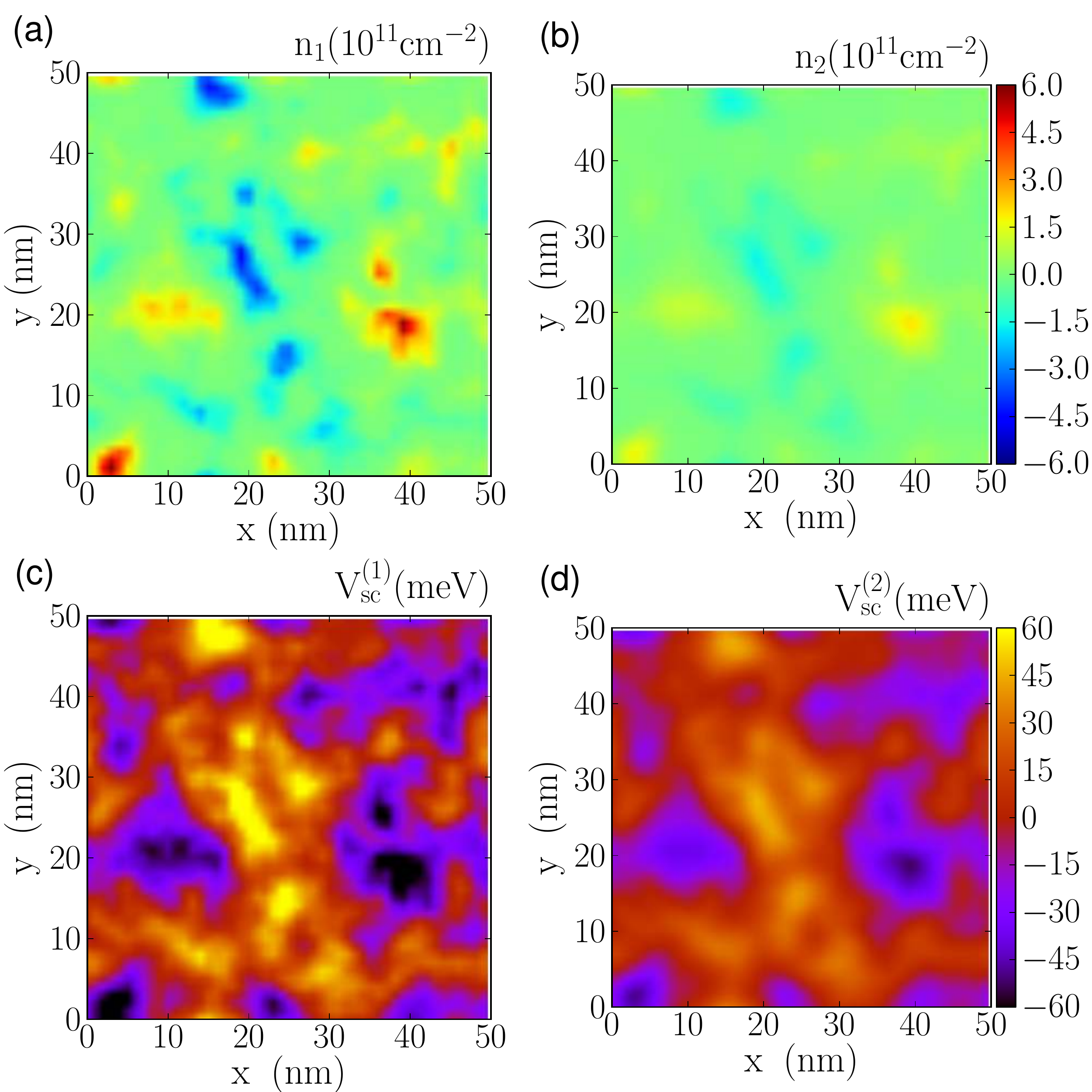}
   \caption{
           (Color online). Color plots showing (a) $\rm{n}_1$(\rr), (b) $\rm{n}_2$(\rr), (c) $\rm{V}^{(1)}_{\rm{sc}}(\rr)$, and (d) $\rm{V}^{(2)}_{\rm{sc}}(\rr)$ for a SLG-SLG system at the charge neutrality point for a single disorder realization with $\rm{n}_{\rm{imp}}=3 \times 10^{11} \rm{cm}^{-2}$, $\rm{d} = 1$ nm, and $\rm{d}_{12} = 1$ nm. 
        } 
  \label{fig:figure_2}
 \end{center}
\end{figure}
\begin{figure}[htb]
 \begin{center}
  \centering
   \includegraphics[width=8.5cm]{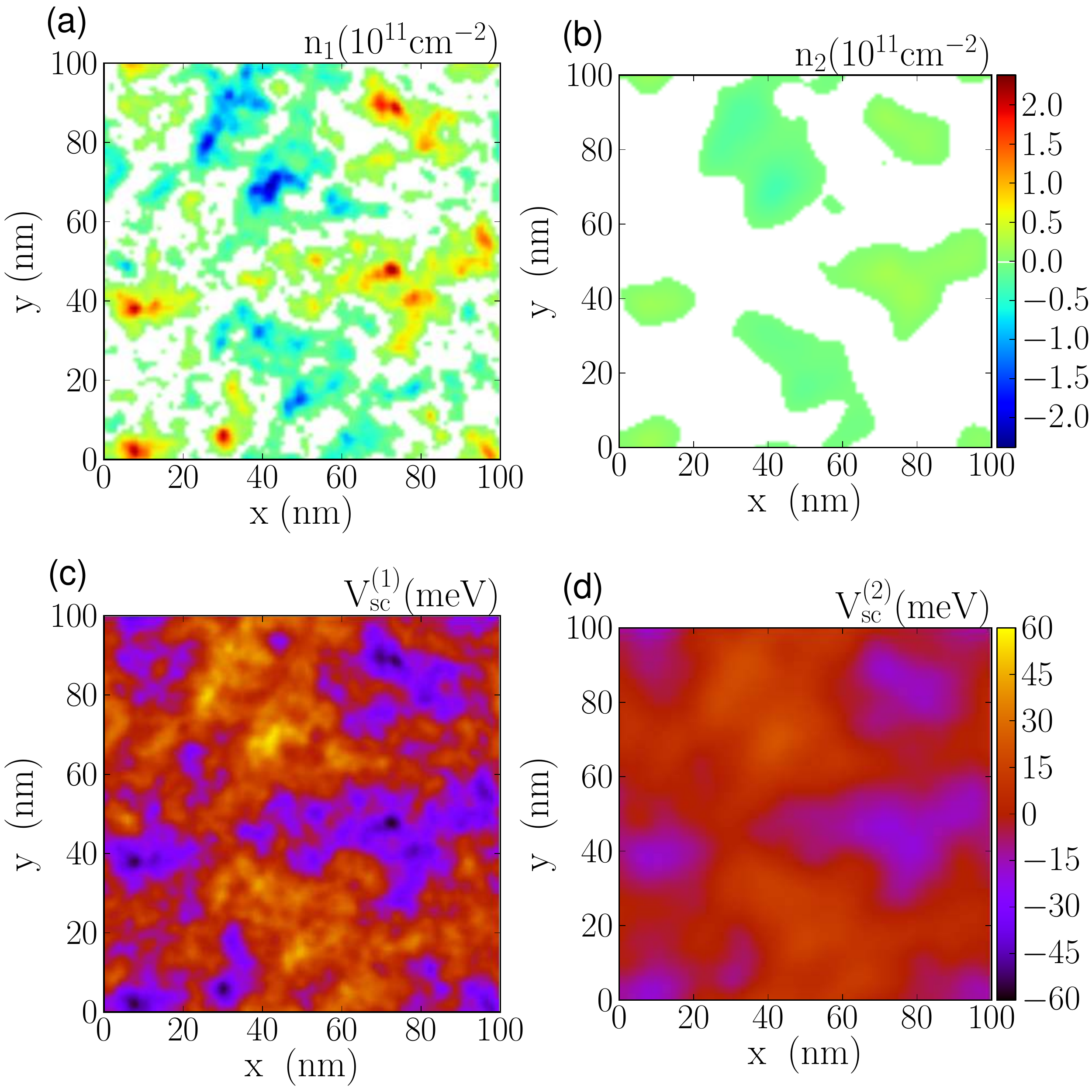}
   \caption{
           (Color online). Color plots showing (a) $\rm{n}_1$(\rr), (b) $\rm{n}_2$(\rr), (c) $\rm{V}^{(1)}_{\rm{sc}}(\rr)$, and (d) $\rm{V}^{(2)}_{\rm{sc}}(\rr)$ for a SLG-SLG system
 at the charge neutrality point for a single disorder realization with $\rm{n}_{\rm{imp}}=3 \times 10^{11} \rm{cm}^{-2}$, $\rm{d} = 1$ nm, $\rm{d}_{12} = 1$ nm, and a finite
           band-gap $\Delta=20$~meV in both layers.
        } 
  \label{fig:figure_2b}
 \end{center}
\end{figure}

 
Figure~\ref{fig:figure_2} shows the profiles for a single disorder realization of the carrier density and of the screened
disorder potential in each layer of a SLG-SLG heterostructure, at the neutrality point.
We see that, as for the case of isolated SLG and BLG
\cite{martin2008,rossi2008,zhang2009,deshpande2009,deshpande2009b,dassarma2011},
the carrier density profile breaks up in electron-hole puddles. 
We also notice that the amplitude of the density fluctuations and the strength of the screened
disorder potential in the top layer
is much smaller than in the bottom layer. This is due mostly to the screening
of the charge impurities by the layer closer to the impurities.
When the spectrum of SLG is gapped some regions of the samples will be insulating.
This is shown by Fig.~\ref{fig:figure_2b} which presents the density and screened disorder
profiles for a single disorder realization in a SLG-SLG system in which the 
band-gap in both graphene layers is set equal to 20~meV.
The white areas in Fig.~\ref{fig:figure_2b}~(a),~(b) are insulating regions, 
i.e. regions in which the local chemical potential is within the band-gap and
therefore contain no carriers.
The results shown in Figs.~\ref{fig:figure_2},\ref{fig:figure_2b} show how the profiles of the density and disorder of the top layer and the bottom layer are different. The asymmetry between the profiles in the two layers will also be reflected in the transport properties as observed experimentally
\cite{ponomarenko2011}. In particular, for our configuration in which the disorder is dominated
by the charge impurities at the surface of the \sio, we see that in the presence of a gap the 
insulating regions are substantially larger in the top layer than in the bottom layer.
We discuss the effect of this asymmetry on the qualitative features of electronic transport
in section~\ref{sec:mit}.

\begin{figure}[htb]
 \begin{center}
  \centering
   \includegraphics[width=8.5cm]{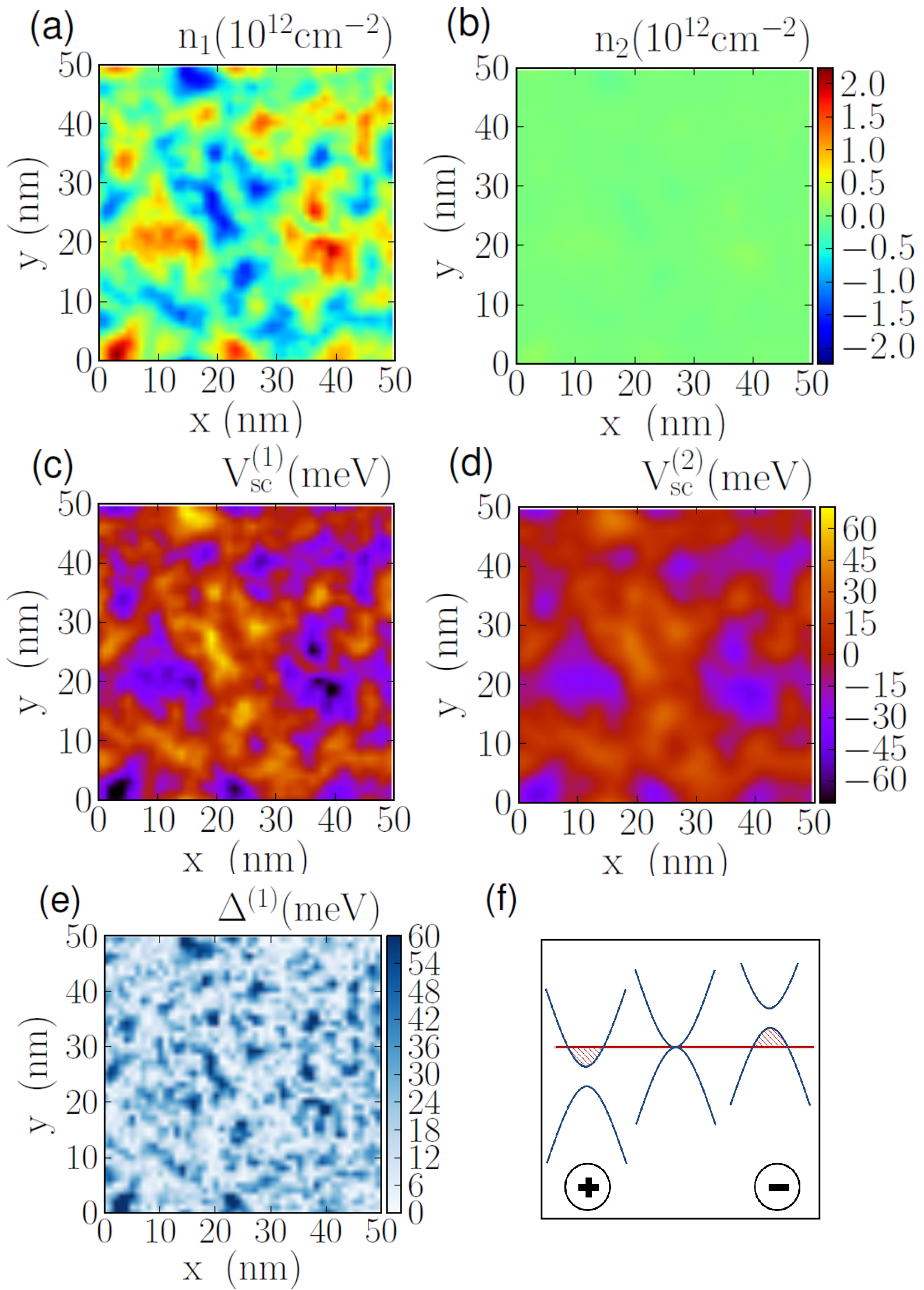}
   \caption{
           (Color online). Color plots showing (a) $\rm{n}_1$(\rr), (b) $\rm{n}_2$(\rr), (c) $\rm{V}^{(1)}_{\rm{sc}}(\rr)$, (d) $\rm{V}^{(2)}_{\rm{sc}}(\rr)$ and (e) $\Delta^{(1)}$ corresponding to the BLG-SLG hybrid system at charge neutrality point for a single disorder realization, $\rm{n}_{\rm{imp}}=3 \times 10^{11} \rm{cm}^{-2}$, $\rm{d} = 1$ nm, and $\rm{d}_{12} = 1$ nm. (f) Sketch of the gapped BLG bands in the presence of disorder.  
        } 
  \label{fig:figure_3}
 \end{center}
\end{figure}
Figure~\ref{fig:figure_3} shows the profiles for a single disorder realization
of the carrier density, panels (a) and (b),  
and screened disorder potential, panels (c) and (d), in each layer
of a hybrid BLG-SLG heterostructure at the charge neutrality point.
In comparing Fig.~\ref{fig:figure_2}~(a) and Fig.~\ref{fig:figure_3}~(a), 
we notice that
the carrier density inhomogeneities are much stronger for BLG than SLG
(all the rest being the same). 
This is due to the difference in the low-energy band structure
between SLG and BLG.
Due to this difference, the price in kinetic energy to create a density fluctuation
at low energies is much higher for SLG than BLG. 
%
Figure~\ref{fig:figure_3}~(b) shows that the amplitude of the density fluctuations in the top layer (SLG)
is much smaller in BLG-SLG than in the SLG-SLG.
This is due to the fact that BLG, as the layer closer to the impurities,
is much more efficient than SLG in screening the second layer from
the disorder potential due to the charge impurities.
This indicates that the mobility of SLG could be increased significantly
when placed in a heterostructure in which the layer closest to the 
charge impurities is BLG. That this is the case is further confirmed 
by the disorder-averaged results that we present below.

Figure~\ref{fig:figure_3}~(e) shows the profile for single disorder realization of the band-gap in BLG.
We see that, due to the presence of the charge impurities, $\Delta$ is very
inhomogenous. In addition, we see that locally $\Delta$ can be as large as 60~meV.
One could then wonder why in correspondence with the regions where $\Delta$
is large, the carrier density, Fig.~\ref{fig:figure_3}~(a), locally does not go to zero.
%
%
This is due to the fact that when the doping is set to zero
in both layers the perpendicular electric field responsible
for opening the band-gap is due to the charge impurities that
we have assumed to be concentrated below the first layer.
In these conditions, the regions in which $E_\perp$ is strong
correspond to regions where the density of charge impurities
is high and the induced carrier density is also high.
In other words, for the conditions considered, regions
where $\Delta\neq 0$ are also regions where the local
value of the chemical potential is outside the gap 
as shown schematically in Fig.~\ref{fig:figure_3}~(f).
The scenario sketched in Fig.~\ref{fig:figure_3}~(f)
is not valid when a non-negligible  density
of charge impurities is also present above the top graphenic
layers or between the two graphenic layers.
Also, when the doping in one or both the two graphenic layers is not zero
there will be a uniform contribution to $E_\perp$ 
and this can create regions where the chemical potential
is within the gap.
\begin{figure}[htb]
 \begin{center}
  \centering
   \includegraphics[width=8.5cm]{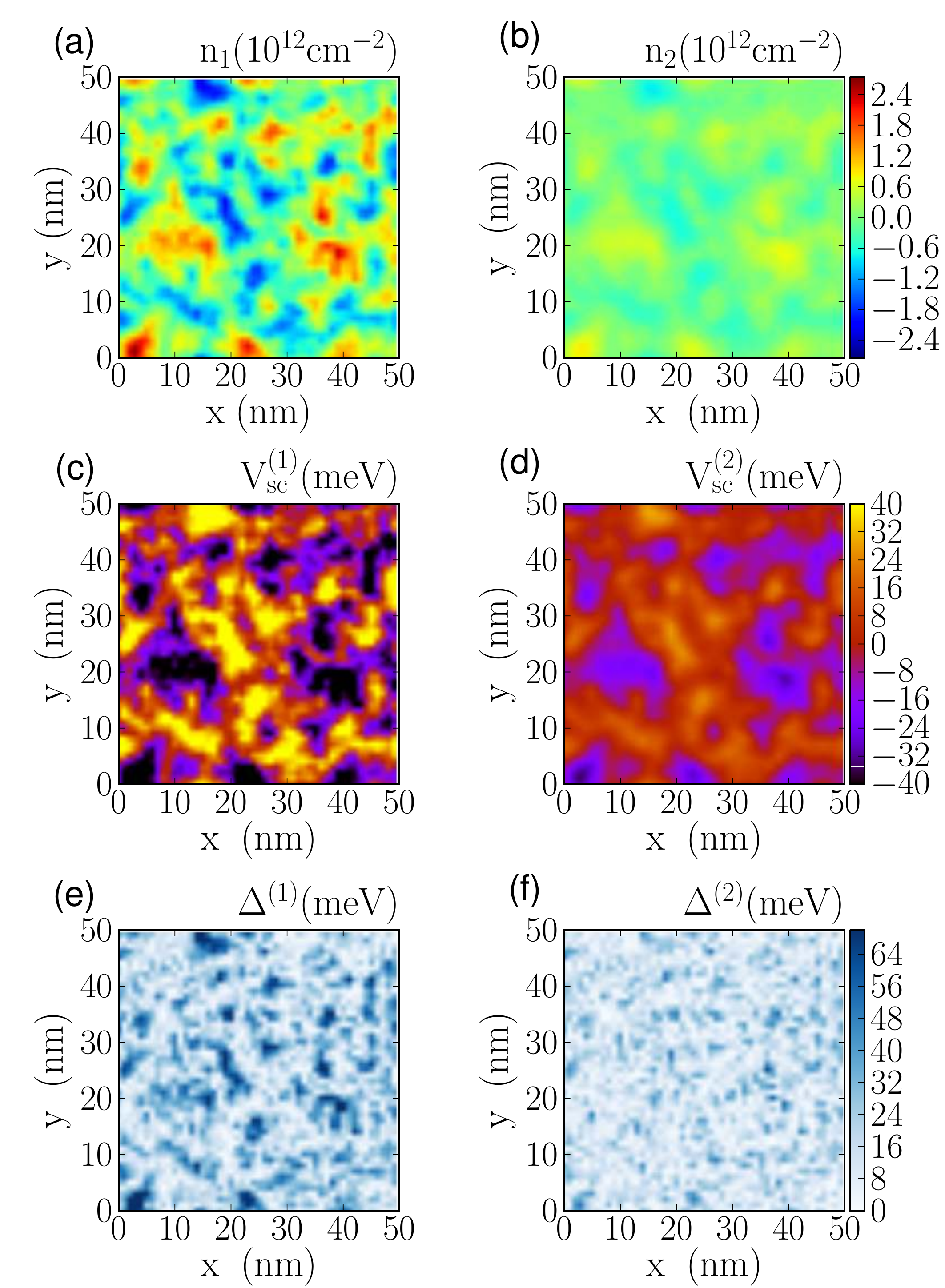}
   \caption{
           (Color online). Color plots showing (a) $\rm{n}_1$(\rr), (b) $\rm{n}_2$(\rr), (c) $\rm{V}^{(1)}_{\rm{sc}}$, (d) $\rm{V}^{(2)}_{\rm{sc}}$, (e) $\Delta^{(1)}$, and (f) $\Delta^{(2)}$ corresponding to the BLG-BLG system at charge neutrality point for a single disorder realization, $\rm{n}_{\rm{imp}}=3 \times 10^{11} \rm{cm}^{-2}$, $\rm{d} = 1$ nm, and $\rm{d}_{12} = 1$ nm. 
        } 
  \label{fig:figure_4}
 \end{center}
\end{figure} 
Figure~\ref{fig:figure_4} shows the profiles for single disorder realization
of carrier density, screened disorder potential, and gap, in both layers
of a BLG-BLG heterostructure, at the neutrality point.
As for the other heterostructures, we see that the screening
by the first layer considerably reduces the amplitude of the density
inhomogeneities in the second layer and of the screened disorder potential.
In addition, the band gap in the second layer is quite smaller than that in the first layer as we see in Fig.~\ref{fig:figure_4}~(e), (f).

A quantitative comparison between the theoretical and the experimental
results is only possible by obtaining the disorder-averaged values of
the quantities that are measured experimentally. In addition, the
disorder-averaged characterization of the ground state carrier
density distribution is an essential ingredient for the development
of the transport theory in the presence of strong, disorder-induced, 
carrier density inhomogeneities
\cite{dassarma2011}.

\begin{figure}[!th]
 \begin{center}
  \centering
  \includegraphics[width=8.5cm]{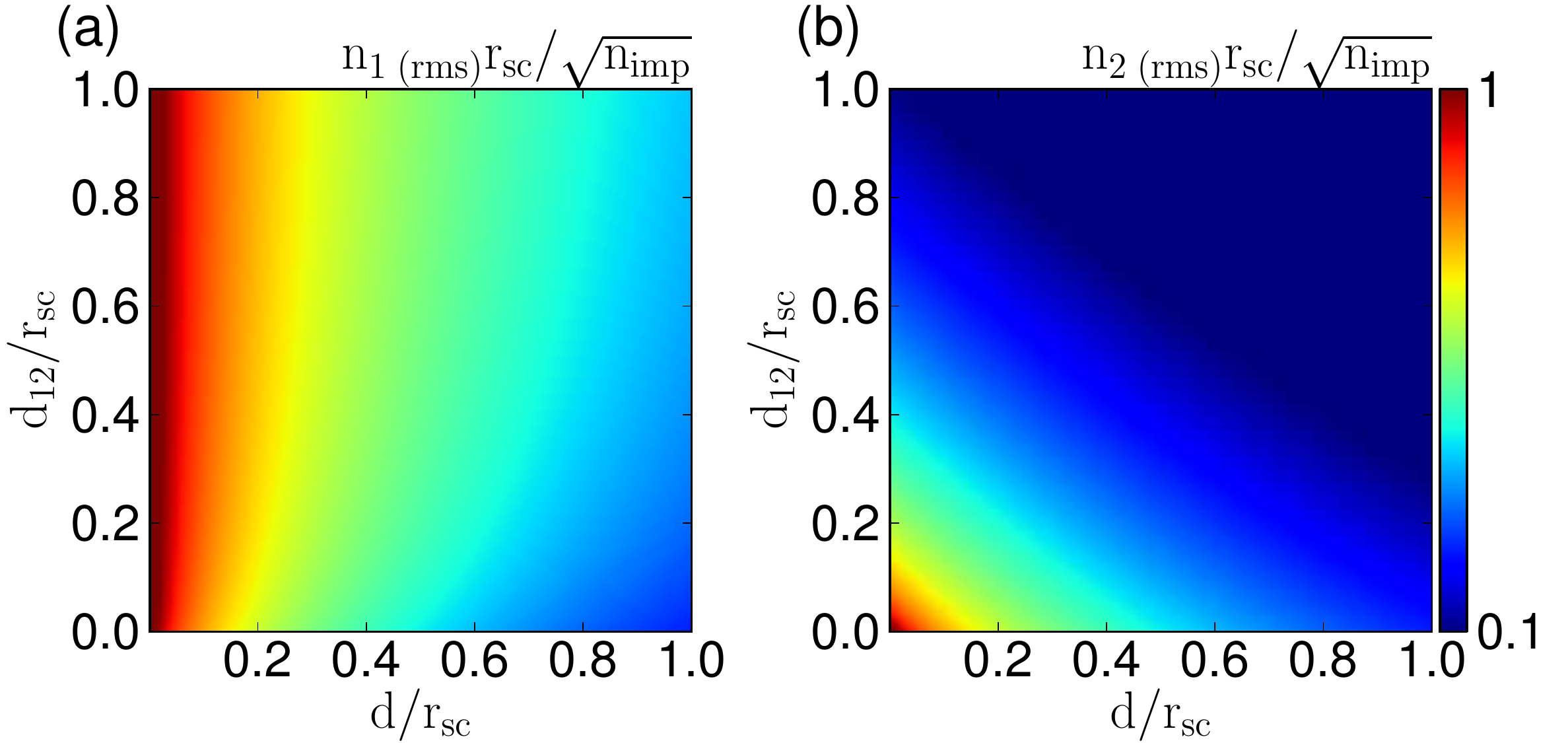}
  \caption{
           (Color online). 
           Color plots of (a) $\rm n_{1 \; (rms)} r_{sc} / \sqrt{\rm n_{imp}} $, and (b) $\rm n_{2 \; (rms)} r_{sc} / \sqrt{\rm n_{imp}}$ as a function of $\mbox{d}/\rb$, and $\mbox{d}_{12}/\rb$ as obtained in equation \ref{eq:nrms-gapless-blg}.
         } 
  \label{fig:integrals_vs_dd12}
 \end{center}
\end{figure}

For BLG-BLG heterostructures in the limit in which the band-gap $\Delta$ is zero,
we can obtain analytic expressions for
the disorder-averaged quantities that characterize the density profile and
the screened disorder potential from the TFDT equations. 
Below we will show that in some situations the results obtained by 
setting $\Delta=0$ provide results for for $\nrms$ and $\vrms$
that well approximate the results obtained by calculating $\Delta$
self-consistently.
%
%
By minimizing the functional $E[n_1, n_2]$ of BLG-BLG structures with $\Delta=0$ with respect to the density
profile $n_1(\rr)$ in the first layer and the density profile $n_2(\rr)$ in the second layer
we find:
\begin{align}
 n_i (\qq)  =  &\frac{ \rb |\qq| e^{|\qq|d_{12}}}{\pi \left[e^{2|\qq|d_{12}}\left(1+|\qq| \rb \right)^2 -1 \right]} \times \nonumber\\
               &\left[\frac{V_D^{(j)}(\qq)}{r_{sc}} - 
                \frac{2 m^*}{\hbar^2} \mu_j \delta(\qq)\right. + \nonumber \\
               &\left.e^{|\qq|d_{12}}\left(1+|\qq| \rb\right) \left(\frac{2 m^*}{\hbar^2}\mu_i\delta(\qq)- \frac{V_D^{(i)}(\qq)}{\rb} \right)\right] 
 \label{eq:ndblg}
\end{align}
%
%
%
%
where $n_i(\qq)$ is the Fourier transform of the carrier density profile in layer $i=1,2$, 
$j=2 \;(1)$ if $i=1\;(2)$, and 
$\rb=\epsilon \hbar^2/(2e^2m^*) \approx 3.2$~nm is the BLG screening length.
Using the statistical properties of the impurity distribution $c(\rr)$
we can calculate the root mean square of the carrier densities ($\rm{n}_{i(\rm{rms})}$) and the screened disorder potential
\begin{align*}
V^{(i)}_{sc} &= \frac{V_D^{(i)}(\rr)}{\rb} + \frac{1}{2 \rb}\int d\rr'\frac{n_j(\rr')}{\left[|\rr-\rr'|^2+d_{12}^2 \right]^{1/2}} \\ 
             & + \frac{1}{2 \rb} \int d\rr' \frac{n_i(\rr')}{|\rr-\rr'|}.
\end{align*}
We find:
\begin{align}
\rm{n}_{i(\rm{rms})}&=\left[\frac{2}{\rb^2\pi} \nimp \mbox{I}_i \left(\frac{d}{\rb},\frac{d_{12}}{\rb} \right) \right]^{1/2},
 \label{eq:nrms-gapless-blg}\\
V^{(i)}_{\rm sc({rms})} &= \frac{\hbar^2 \pi}{2m^*} \rm{n}_{i(\rm{rms})},
 \label{eq:vrms-gapless-blg}
\end{align}
($i=1,2$) where 
\beq
 \mbox{I}_1 (x,y) = \int_0^{\infty} dz z  e^{-2xz} \frac{\left[1-e^{2yz}(1+z) \right]^2}{ \left[1-e^{2yz}(1+z)^2  \right]^2 }, 
 \label{eqn:int1}
\enq
and
\beq
 \mbox{I}_2 (x,y) = \int_0^{\infty} dz \;\; \frac{ z^3 e^{2z(y-x)} }{ \left[1-e^{2yz}(1+z)^2  \right]^2 }
 \label{eqn:int2} \; .
\enq
Figure~\ref{fig:integrals_vs_dd12} shows the scaling of $\nrms$ (and $\vrms$) in the two layers
as a function of $d/\rb$ and $d_{12}/\rb$. As $d$ increases the amplitude of the carrier density
inhomogeneities decreases rapidly. 
As $d_{12}$ increases, $\nrmsone$ approaches the value found for a single BLG sheet 
\cite{rossi2011} whereas $\nrmstwo$ decreases exponentially to zero. 

As discussed in Sec.~\ref{sec:method} when SLG is one of the constituents of
the heterostructure, and/or when the BLG's band-gap cannot be neglected,
the TFDT equations can only be solved numerically due to the nonlinearity 
induced by the kinetic energy term. Below we present our results for the
disorder-averaged quantities. Apart when explicitly indicated,
all the results were obtained for $160\times 160$~nm samples
with a spatial coarse-graining of 1~nm
\cite{shklovskii1984,efros1993}.
For each case we used a number of disorder realizations, $N_S$,
large enough to guarantee that the results would not change
if a larger number of disorder realizations were used.
For the cases presented below we find that 
the results do not depend on $N_s$ when $N_s$ is larger
than 500. 

\begin{figure}[htb]
 \begin{center}
  \centering
   \includegraphics[width=8.5cm]{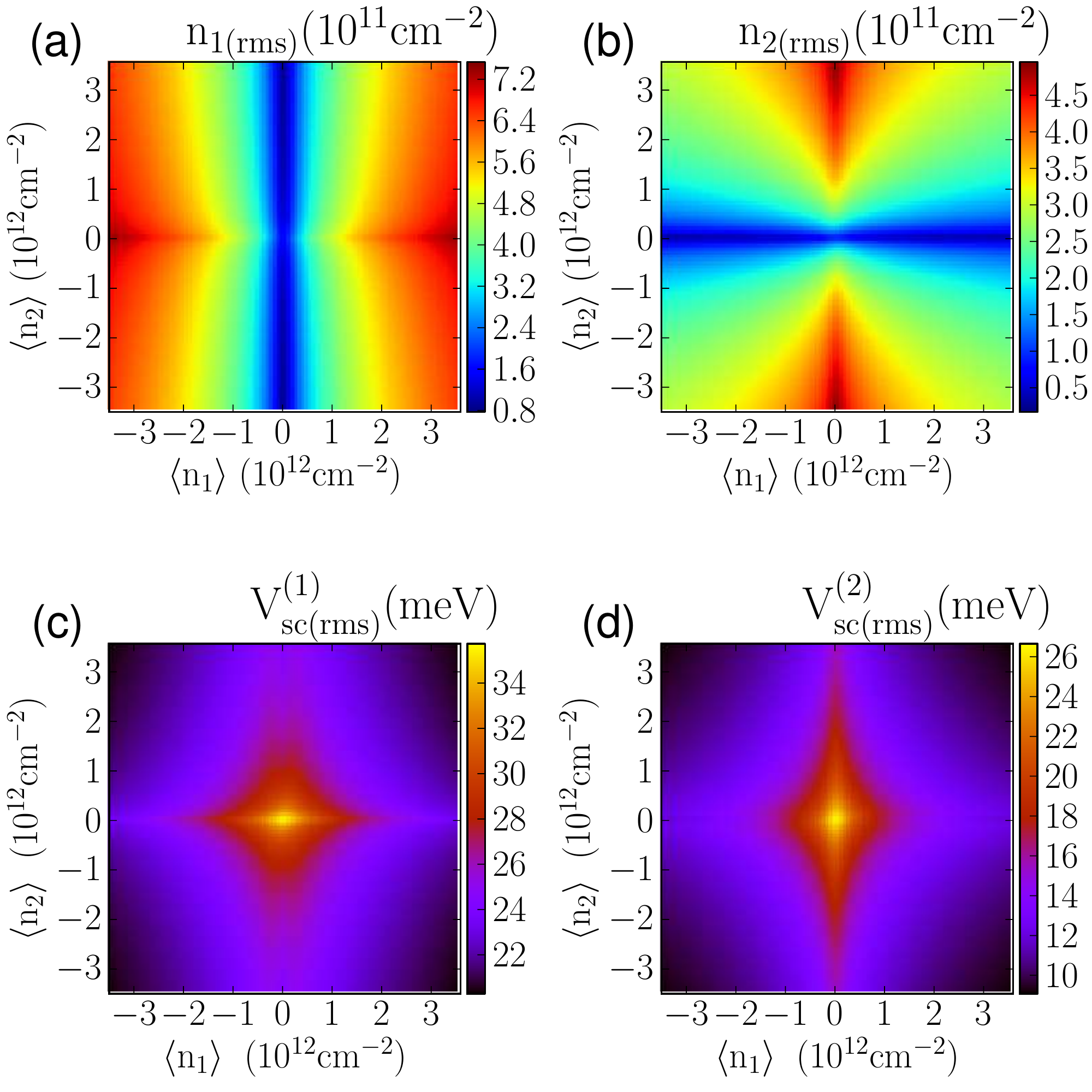}
   \caption{
           (Color online). Color plots of (a) $\rm{n}_{1 \; (\rm{rms})}$, (b) $\rm{n}_{2 \; (\rm{rms})}$, (c) $\rm{V}^{(1)}_{sc \; (\rm{rms})}$, and  (d) $\rm{V}^{(2)}_{sc \; (\rm{rms})}$ for SLG-SLG system as a function of the average carrier density for $\rm{n}_{\rm{imp}}=3 \times 10^{11} \rm{cm}^{-2}$, $\rm{d} = 1$ nm, and $\rm{d}_{12} = 1$ nm.
        } 
  \label{fig:figure_6}
 \end{center}
\end{figure} 
Figure~\ref{fig:figure_6} shows the root mean square of the carrier
density and of the screened disorder potential in each layer of
a SLG-SLG heterostructure. 
We see that the amplitude of the carrier
density fluctuations in the first layer increases with 
$\nava$ and depends quite weakly on $\navb$.
Analogously, $\nrms$ in the second layer increases with $\navb$.
This is due to the fact that as the doping increases more
carriers are available to screen the disorder potential by creating
high density electron (hole) puddles in correspondance of the valleys (peaks)
of the bare disorder potential. 
However, we see that $\nrmsb$ also depends significantly on 
$\nava$. This is due to the fact that the first layer, being the
closest to the charge impurities, is most responsible for the
screening of the disorder potential and therefore significantly
affects the amplitude of the density fluctuations in the second layer.
Both $\nava$ and $\navb$ contribute to a decrease of the screened
disorder potential in layer 1 and layer 2, as show by Fig.~\ref{fig:figure_6}~(c) and (d).
The results of Fig.~\ref{fig:figure_6}~(b) and (d) confirm the conclusion
that we derived from the single disorder realization results: due to 
the screening effect of the first layer the amplitude of the carrier
density inhomogeneities and the strength of the screened disorder potential
are weaker in layer 2 than in layer 1.

\begin{figure}[htb]
 \begin{center}
  \centering
   \includegraphics[width=8.5cm]{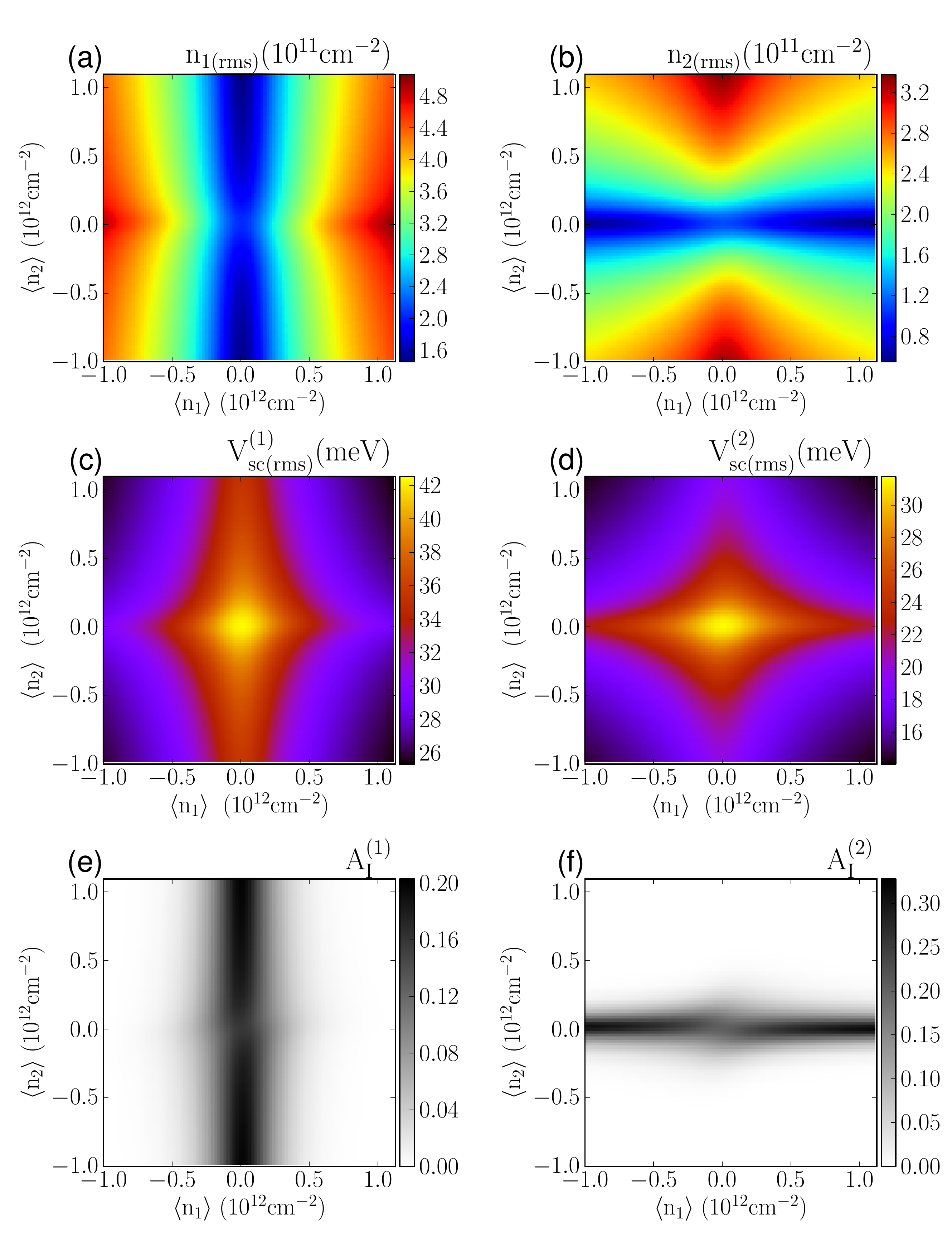}
   \caption{
           (Color online). Color plots of (a) $\rm{n}_{1 \; (\rm{rms})}$, (b) $\rm{n}_{2 \; (\rm{rms})}$, (c) $\rm{V}^{(1)}_{sc \; (\rm{rms})}$, (d) $\rm{V}^{(2)}_{sc \; (\rm{rms})}$,
           (e) fraction of the area of the sample that is insulating in layer 1, $A_I^{(1)}$, and
           (f) fraction of the area of the sample that is insulating in layer 2, $A_I^{(2)}$,
           for SLG-SLG system with finite band-gap as a function of the average carrier density for 
           $\Delta=20$~meV, $\rm{n}_{\rm{imp}}=3 \times 10^{11} \rm{cm}^{-2}$, $\rm{d} = 1$ nm, and $\rm{d}_{12} = 1$ nm.
        } 
  \label{fig:figure_6b}
 \end{center}
\end{figure}
\begin{figure}[!!!!!!!t!!!!!!!!!!!]
 \begin{center}
  \centering
   \includegraphics[width=8.5cm]{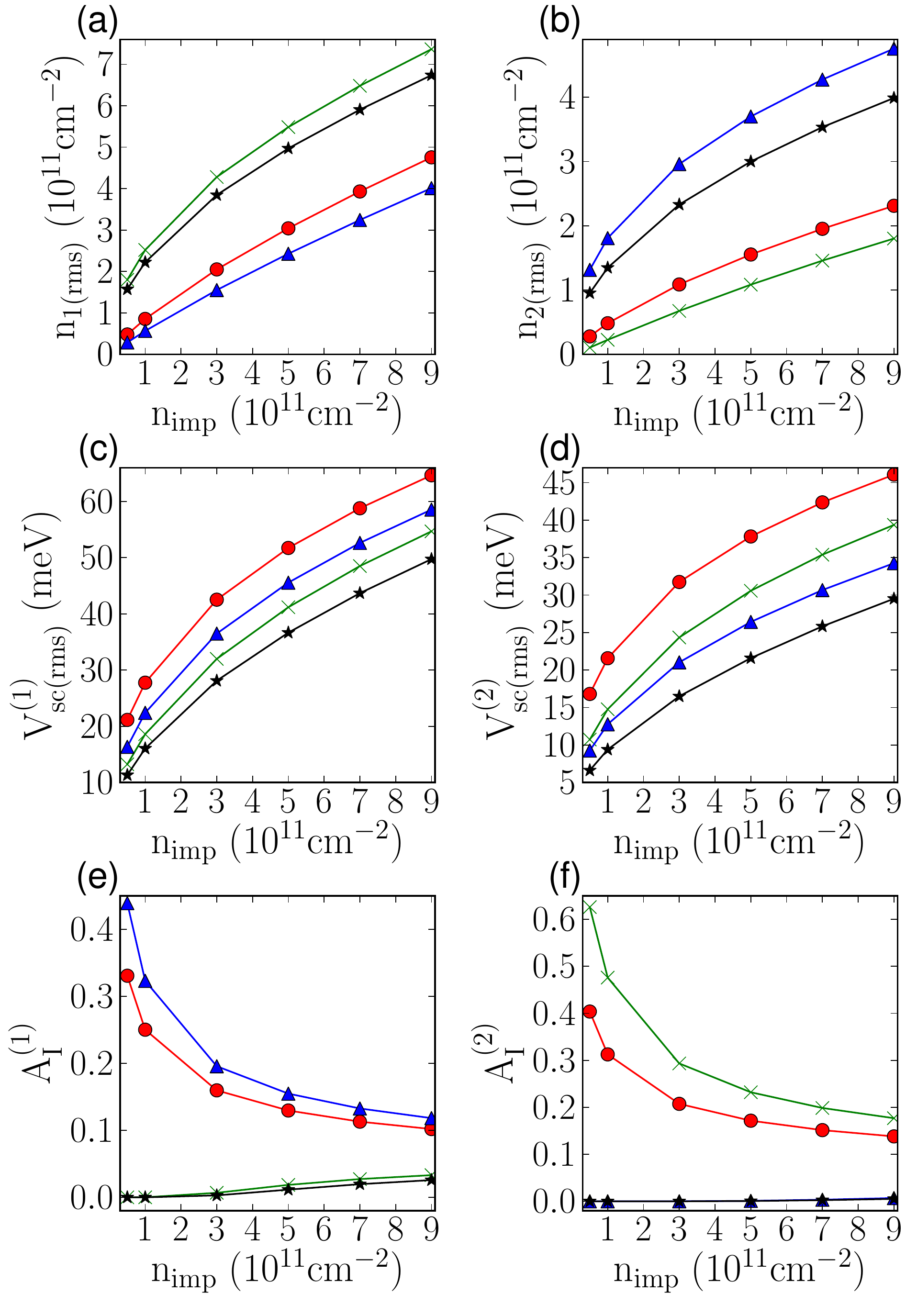}
   \caption{
           (Color online). Plots of (a) $\rm{n}_{1 \; (\rm{rms})}$, (b) $\rm{n}_{2 \; (\rm{rms})}$, (c) $\rm{V}^{(1)}_{sc \; (\rm{rms})}$, (d) $\rm{V}^{(2)}_{sc \; (\rm{rms})}$, (e) fraction of the area of the sample that is insulating in layer 1, $A_I^{(1)}$, and (f) fraction of the area of the sample that is insulating in layer 2, $A_I^{(2)}$,  as a function of the impurity strength $\nimp$ for a SLG-SLG system with gapped graphene: $\Delta=20$~meV, d$ = 1 $ nm, d$_{12} = 1$ nm, and for four different carrier density averages. The circle symbols correspond to $\langle \rm{n}_1 \rangle = 0$ cm$^{-2}$ and $\langle \rm{n}_2 \rangle = 0$ cm$^{-2}$, the cross symbols to $\langle \rm{n}_1 \rangle = 5\times 10^{11}$ cm$^{-2}$ and $\langle \rm{n}_2 \rangle = 0$ cm$^{-2}$, the triangle symbols to $\langle \rm{n}_1 \rangle = 0$ cm$^{-2}$ and $\langle \rm{n}_2 \rangle = 5\times 10^{11}$ cm$^{-2}$, and the star symbols correspond to $\langle \rm{n}_1 \rangle =  5\times 10^{11}$ cm$^{-2}$ and $\langle \rm{n}_2 \rangle = 5\times 10^{11}$ cm$^{-2}$.
        } 
  \label{fig:figure_12_b}
 \end{center}
\end{figure}

 
In presence of a band-gap in the graphene spectrum for SLG-SLG systems, the dependence of $\nrms$ and $\vrms$ on $\nava$ and $\navb$
is qualitatively similar to the gapless cases. In the presence of a gap it is interesting to also look at how the fraction of the area
of graphene that is insulating, $A_I^{(1)}$ ($A_I^{(2)}$) for layer 1 (2), depends on the doping in the two layers, see Figs.~\ref{fig:figure_6b}~(e),~(f).
For relatively large impurity densities, such as considered for the results shown in Fig.~\ref{fig:figure_6b}~(e),~(f), $A_I$ in layer 1 depend
only weakly on the doping of layer 2, and vice versa. However, as we show in Fig.~\ref{fig:figure_12_b}, and as we discuss in section~\ref{sec:mit}, this is not the case
at low impurity densities. In practice we have that when the screened disorder $\vrms\lesssim\Delta$ the effect of layer
$j$ on $A_I$ of the other layer can be very significant.

\begin{figure}[!!!!!t!!!!]
 \begin{center}
  \centering
   \includegraphics[width=8.5cm]{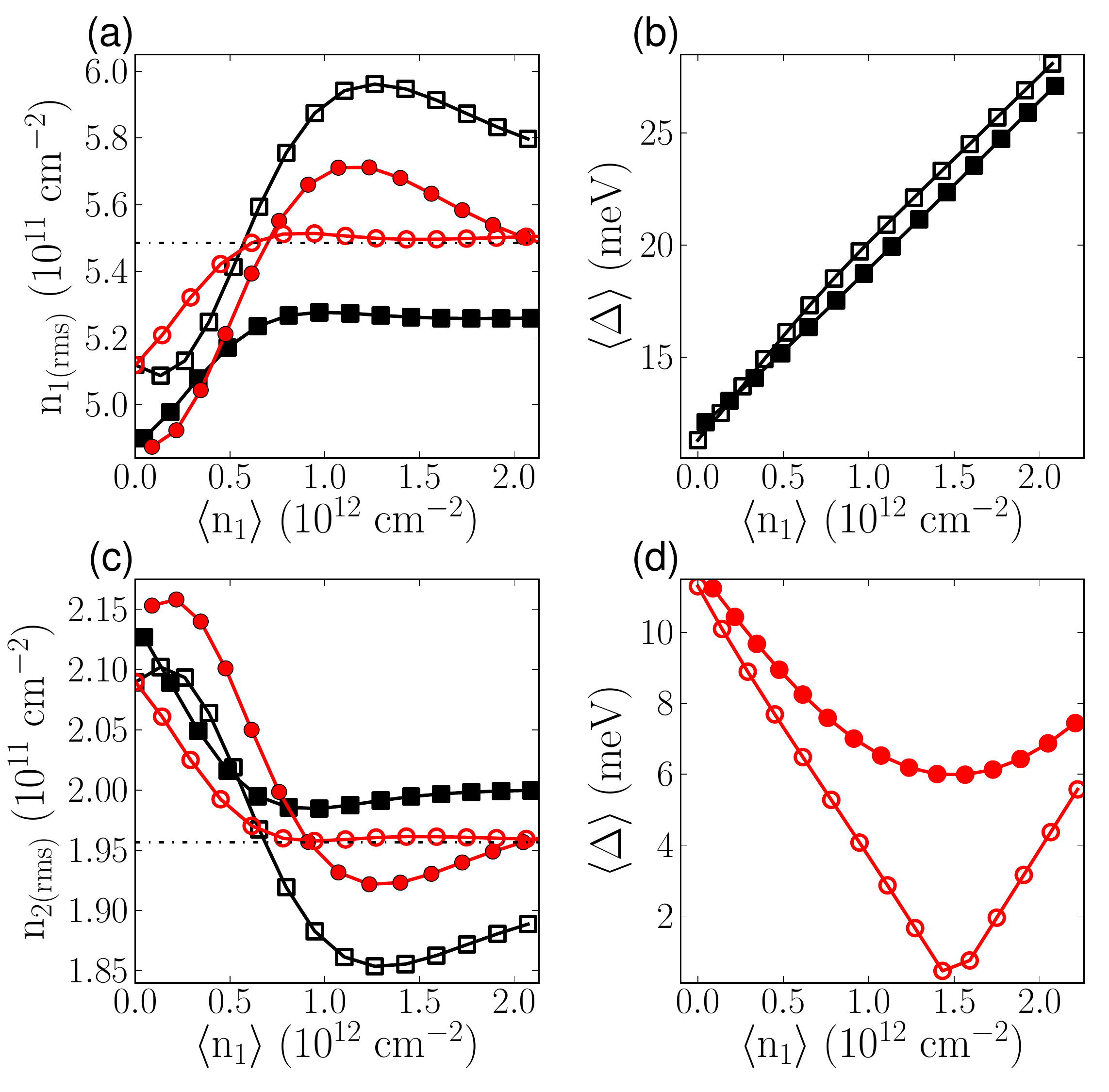}
   \caption{
           (Color online). Plot of (a) $\nrmsa$ and (c) $\nrmsb$ as a function of $\nava$ for $\nimp=2 \times 10^{11} \mbox{cm}^{-2}$, d$_{12}=1$nm, and d$=1$nm. The squares symbols correspond to $\langle \mbox{n}_2 \rangle = 1.5 \times 10^{12} \mbox{cm}^{-2}$, and the circle symbols correspond to $\langle \mbox{n}_2 \rangle = -1.5 \times 10^{12} \mbox{cm}^{-2}$. The curves with open symbols show the results obtained keeping $\Delta$ fixed, whereas the curves with solid symbols show the results obtained by calculating $\Delta$ self-consistently. $\langle \Delta \rangle $ is shown in subplots (b) and (d) also as a function of $\nava$. The dashed lines correspond to the case $\Delta = 0$eV for both values of $\navb$, since the gapless BLG-SLG system is even in $\navb$.
        } 
  \label{fig:figure_7}
 \end{center}
\end{figure} 
%
%

%
\begin{figure}[!!!!!t!!!!!]
 \begin{center}
  \centering
   \includegraphics[width=8.5cm]{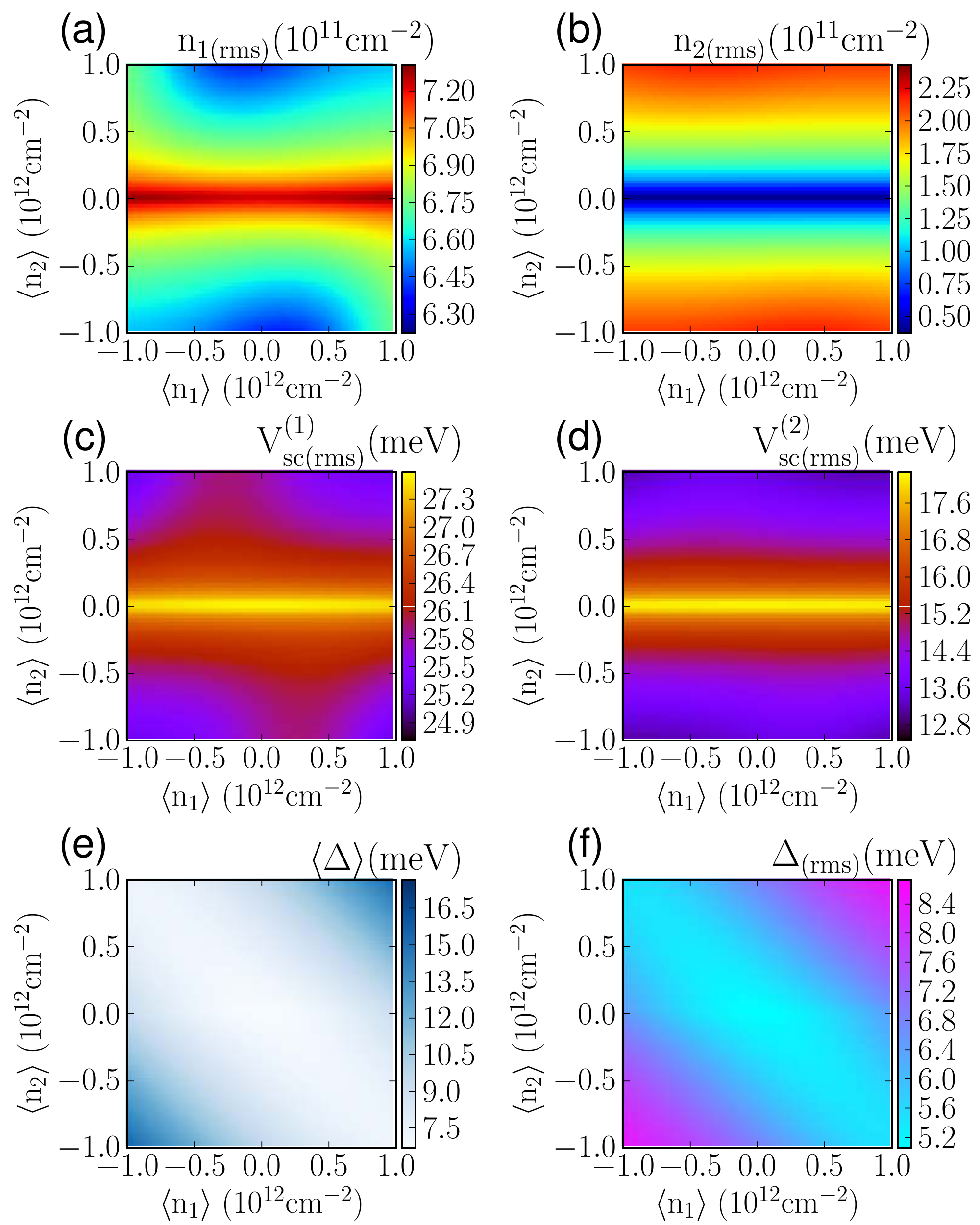}
   \caption{
           (Color online). Color plots of (a) $\rm{n}_{1 \; (\rm{rms})}$, (b) $\rm{n}_{2 \; (\rm{rms})}$, (c) $\rm{V}^{(1)}_{sc \; (\rm{rms})}$, (d) $\rm{V}^{(2)}_{sc \; (\rm{rms})}$, (e) $\langle \Delta \rangle$, and (f) $\Delta_{\rm{rms}}$ for BLG-SLG system as a function of the average carrier density for $\rm{n}_{\rm{imp}}=3 \times 10^{11} \rm{cm}^{-2}$, $\rm{d} = 1$ nm, and $\rm{d}_{12} = 1$ nm.
        } 
  \label{fig:figure_8}
 \end{center}
\end{figure} 

For heterostructures in which BLG is present we need to account for
the opening of a band-gap due to the presence of a perpendicular electric field.
The calculation of the band-gap has to be done self-consistently due
to the fact that the redistribution of the charges in the layer forming
the heterostructure modifies the profile of the perpendicular component
of the electric field, affecting the profile of the band-gap that itself
affects the screening properties of the heterostructure.
%
%
To test the importance of self-consistently calculating the profile of $\Delta$ for a set of cases for BLG-SLG structures,
we first performed the calculation setting $\Delta$ equal to the value obtained
from Eqs.~\ceq{eqn:perp_efield},~\ceq{eqn:potentialenergy},~\ceq{eqn:bandgap} in the limit of homogenous density profiles in the two layers, 
with $n_1=\nava$, and $n_2=\navb$. We then redid the calculation 
by obtaining $\Delta(\rr)$ self-consistently.
The comparison of the two sets of results is shown in 
Fig.~\ref{fig:figure_7} in which $\nrms$ in the two layers and the average gap ($\Deltaav$) are plotted
as a function of $\nava$ for a fixed, non zero, value of $\navb$: the curves with open symbols show
the results obtained keeping $\Delta$ fixed, whereas the curves with solid symbols show
the results obtained by calculating $\Delta$ self-consistently. We see that in general
the value of $\nrms$ obtained using the two approaches differ.
For the case in which $\nava\navb>0$ we have that the value of $\Deltaav$ obtained
self-consistently is reasonably approximated by the fixed value, $\Delta_{\rm fixed}$, obtained assuming
uniform carrier density profiles.
However, for $\nava\navb<0$ we find that the value of $\Deltaav$ is significantly
different from $\Delta_{\rm fixed}$, Fig.~\ref{fig:figure_7}~(d).
The results of Fig.~\ref{fig:figure_7} show that
the effect of the disorder cannot be captured by a simple average of a spatially
homogenous theory and requires a self-consistent calculation of the parameters
defining the local band-structure.
All results that we present for heterostructures in which BLG
is present were obtained calculating $\Delta$ self-consistently.
\begin{figure}[!!!!t!!!!!!]
 \begin{center}
  \centering
   \includegraphics[width=8.5cm]{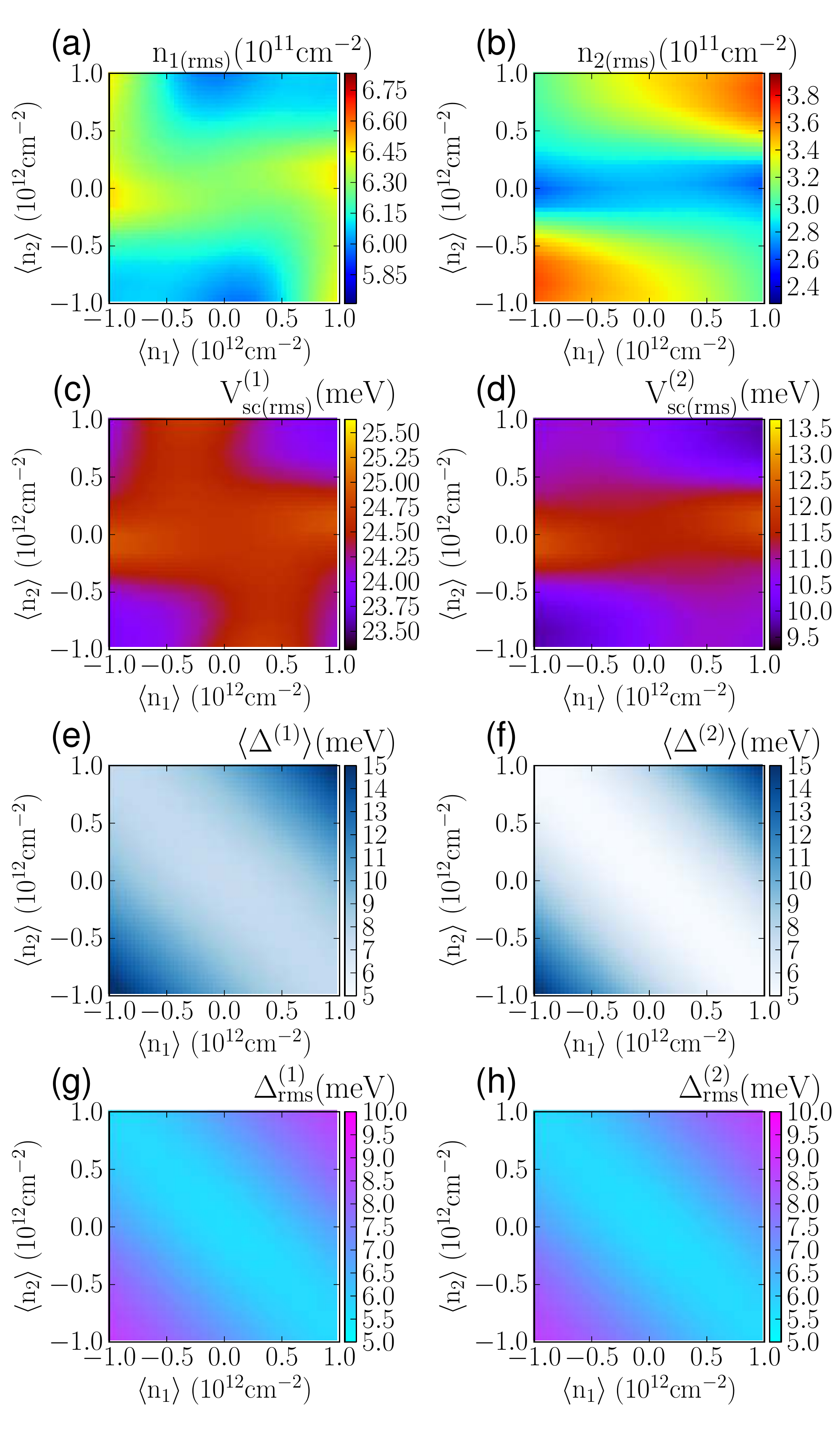}
   \caption{
           (Color online). Color plots of (a) $\rm{n}_{1 \; (\rm{rms})}$, (b) $\rm{n}_{2 \; (\rm{rms})}$, (c) $\rm{V}^{(1)}_{sc \; (\rm{rms})}$, (d) $\rm{V}^{(2)}_{sc \; (\rm{rms})}$, (e) $\langle \Delta^{(1)} \rangle$, (f) $\langle \Delta^{(2)} \rangle$, (g) $\Delta^{(1)}_{\rm{rms}}$, and (h) $\Delta^{(2)}_{\rm{rms}}$ for BLG-BLG system as a function of the average carrier density for $\rm{n}_{\rm{imp}}=3 \times 10^{11} \rm{cm}^{-2}$, $\rm{d} = 1$ nm, and $\rm{d}_{12} = 1$ nm.
        } 
  \label{fig:figure_9}
 \end{center}
\end{figure} 
\begin{figure}[!!!!!!t!!!!!!]
 \begin{center}
  \centering
   \includegraphics[width=8.5cm]{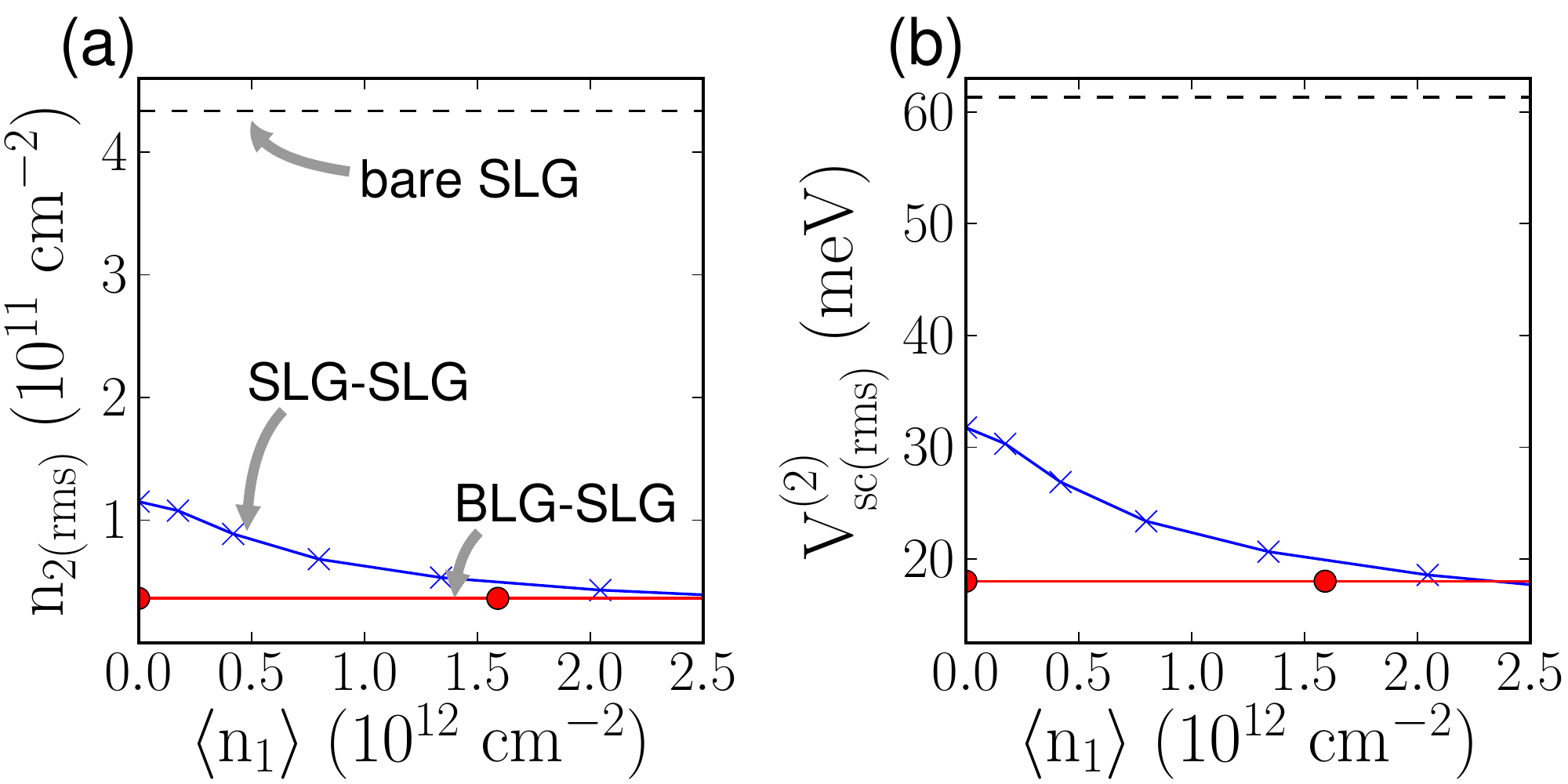}
   \caption{
           (Color online). Plots of (a) $\rm{n}_{2 \; (\rm{rms})}$ and (b) $\rm{V}^{(2)}_{sc \; (\rm{rms})}$ as a function of the carrier density on the graphenic layer closest to the impurities. The blue crosses correspond to the SLG-SLG system, the red circles correspond to the BLG-SLG system, and the black dashed curve correponds to bare SLG.
        } 
  \label{fig:figure_10}
 \end{center}
\end{figure} 

For a fixed $\nimp$, $d$, $d_{12}$,
Fig.~\ref{fig:figure_8} shows the dependence of the disorder averaged quantities characterizing the ground state of a BLG-SLG structure on the $\nava$ and $\navb$.
%
We see that amplitude of the density fluctuations and the strength of the screened disorder
potential at low dopings depend almost exclusively on $\navb$, the average carrier density
in SLG, and only very weakly on $\nava$, the average carrier density in BLG.
This is due to the fact that at low dopings the band gap in BLG is quite small and so
the density of states (DOS) of BLG is to good approximation constant, independent of $\nava$.
On the other hand, in SLG, due to the linear band dispersion, the DOS depends linearly on
the doping $\navb$. As a consequence at low dopings a change of $|\nava|$ has a negligible
effect on the screening properties of the system whereas an increase (decrease) of $|\navb|$ 
increases (decreases) the screening due to the second layer, SLG.
At high dopings the situation is complicated by the effect that a high average density
on each layer has on the size of the gap in BLG, as shown in Fig.~\ref{fig:figure_8}~(e). 
As a consequence the DOS in BLG is
no longer almost independent of $\nava$. This causes dependence of $\nrms$ and $\vrms$
on the value of $\nava$. In particular, the asymmetry of $\nrms$ and $\vrms$
with respect to $\nava$, for large values of $\navb$, is due to the asymmetric
dependence of $\Delta$ on $\nava$, Fig.~\ref{fig:figure_8}~(e).
Figure~\ref{fig:figure_8}~(f) shows the root mean square of $\Delta$, $\Deltarms$.
We see the $\Deltarms$ is in general of the same order of $\Delta$, indicating
the inhomogeneities of the band-gap in BLG are quite strong and cannot be treated
perturbatively. In addition, we see that, qualitatively, $\Deltarms$ depends
on $\nava$ and $\navb$ in a similar way to $\Deltaav$.
Another important feature of the results of Fig.~\ref{fig:figure_8} to notice
is that when both $|\nava|$ and $|\navb|$ are large the size of the gap in BLG
is comparable to the strength of the screened disorder potential.
In these conditions we expect that the transport
properties might be significantly affected by the presence of the band-gap
and that BLG might behave as a bad-metal \cite{rossi2011}.

We now consider the BLG-BLG heterostructure. In this case both the top layer
and the bottom layer can have a gapped band structure. Due to the fact that
the band gap in both layers depends asymmetrically on $\nava$ and $\navb$, Fig.~\ref{fig:figure_9}~(e),~(f),
we find that $\nrms$ and $\vrms$, in both layers, depend asymmetrically
on the average carrier density of each layer, as shown in Fig.~\ref{fig:figure_9}~(a)-(d).
We also find that in both layers the r.m.s. of the band gap is of the same 
order as $\Deltaav$ and that it scales with $\nava$ and $\navb$ qualitatively
as $\Deltaav$.
We notice that for the bottom layer 
the average band-gap
is never larger than the r.m.s of screened disorder potential. On the other hand,
for the top layer we have that at large  $|\nava|$ and $|\navb|$ the average gap
is larger than $\vrmsb$. As a consequence we expect that when $|\nava|$ and $|\navb|$
are large the bottom layer will behave as a bad metal and the top layer as
a bad insulator \cite{rossi2011}.

By comparing the results of Fig.~\ref{fig:figure_6},~\ref{fig:figure_8}, and~\ref{fig:figure_9},
we see that the three heterostructures, SLG-SLG, BLG-SLG, BLG-BLG, exhibit
disorder-induced density fluctuations of comparable magnitude, and comparable
strengths of the screened disorder potential. These results suggest
that the effect of disorder on the establishment of collective ground
states that has been proposed for SLG-SLG 
\cite{min2008b, zhang2008jog, kharitonov2008b, kharitonov2010, zhang2010, jzhang2013}
BLG-SLG \cite{jzhang2013},
and BLG-BLG \cite{perali2013}
should be comparable.

It is interesting to compare the amplitude of $\nrms$ and of $\vrms$
for SLG when isolated and when part, as top layer, of one of the heterostructures considered.
Figure~\ref{fig:figure_10} presents such a comparison. 
As we had anticipated above we see that $\nrms$ and $\vrms$ in SLG are much lower 
when part of a heterostructure, due to the screening of the disorder by the bottom layer,
than when isolated. From the results of Fig.~\ref{fig:figure_10} we see that when the doping in the bottom layer is $\sim 10^{12}{\rm cm}^{-2}$ $\nrms$ can
be reduced by an order of magnitude thanks to the screening of the disorder by the bottom layer.
Figure~\ref{fig:figure_10}~(b) shows that the strength of the screened disorder
potential in SLG is reduced by a factor ~3 by the presence of the graphenic bottom layer.
In addition, Fig.~\ref{fig:figure_10} shows that BLG, as a bottom layer, for
$\nava\lesssim 2.5\times 10^{12}{\rm cm}^{-2}$, is more efficient than SLG to screen 
the top SLG layer. For $\nava\gtrsim 2.5\times 10^{12}{\rm cm}^{-2}$, SLG and BLG,
as bottom layers, have the same effect on screening the disorder for the top
layer given that their band structures are very similar for dopings of this order or larger.

The results of Fig.~\ref{fig:figure_10} suggest that, assuming that charge impurities
are the dominant source of disorder, 
a very effective way to reduce the effects of disorder in SLG and BLG
would be to considerably reduce the thickness of the insulating
layer between the graphene sheet and the back gate. 
Given the modern techniques to realize graphene devices, this is something
that we think could be done using the currently available experimental capabilities.

%
%
%
\begin{figure}[!!!!!t!!!!!!!]
 \begin{center}
  \centering
   \includegraphics[width=8.5cm]{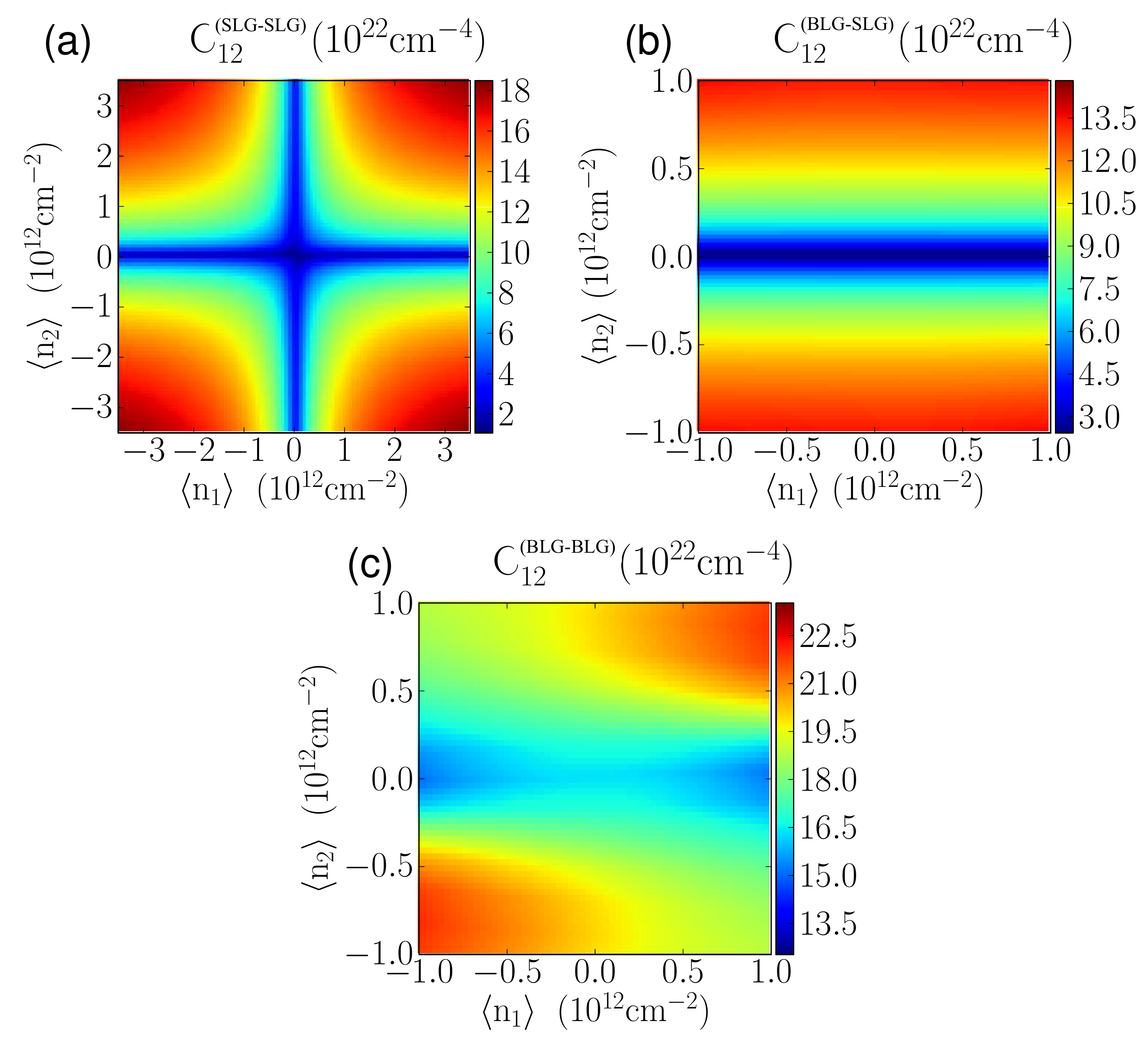}
   \caption{
           (Color online). Color plots of the density correlation $\rm{C}_{12}= \langle \rm{n}_1 \rm{n}_2 \rangle - \langle \rm{n}_1 \rangle \langle \rm{n}_2 \rangle $ as a function of the average carrier density for (a) SLG-SLG,  (b) BLG-SLG and (c) BLG-BLG systems for d$ = 1 $ nm, $\nimp = 3 \times 10^{11} \rm{cm}^{-2}$, and d$_{12} = 1$ nm. 
        } 
  \label{fig:figure_11}
 \end{center}
\end{figure}
%

%
To understand the physics of graphene heterostructures in the presence of disorder a very important property
is the correlation, $C_{12}= \langle \mbox{n}_1(\rr) \mbox{n}_2(\rr)\rangle - \nava \navb$, 
between the density profiles in the two layers. The knowledge of $C_{12}$ is important to
estimate the effect of disorder on the establishment of correlated ground states. 
Moreover, knowledge of the nature of the correlations in the presence of disorder between $\rm{n}_1(\rr)$ and $\rm{n}_2(\rr)$ might be essential to understand recent 
drag resistance measurements 
\cite{gorbachev2012} on SLG-SLG heterostructures.

One possible explanation of these measurements relies on the presence 
of correlated electron hole puddles in the two layers \cite{song2012,song2013} 
close to the double charge neutrality point (i.e. when both $\nava$ and 
$\navb$ are equal to zero). Our results for $C_{12}$, Fig.~\ref{fig:figure_11}, 
show that, for all three heterostructures considered, $C_{12}$ is always positive,
indicating that each electron (hole) puddle in the bottom
layer corresponds an electron (hole) puddle in the top layer.
This is due to the fact that the formation of the electron hole
puddles is mainly due to the presence of charge impurities
below the bottom layer.
%
%
Assuming that the energy transfer mechanism
presented in Ref.~\onlinecite{song2012,song2013} is the main mechanism
for the strong peak of the drag resistivity  observed in Ref.~\onlinecite{gorbachev2012}
at the double charge neutrality point, our results
therefore strongly suggest that in the SLG-SLG double layer structure
used in Ref.~\onlinecite{gorbachev2012}, charge impurities below the bottom layer
are the dominant source of disorder and the main reason for the formation
of the electron-hole puddles at low dopings.

%
If the density of charge impurities between the two graphene sheets, or above
the top sheet, is comparable to the density of charge impurities located
below the bottom sheet the results for the correlation $C_{12}$
would be modified. The amount of change would depend on the details of the device:
ratio between the inter-sheet impurity density, the impurity density above
the top layer,  and the impurity density
below the bottom sheet; average distance of the impurity distributions
to each of the sheets; doping level in each layer,...
In general, we would expect that, as the impurity density between the sheets and above the 
top sheet become less negligible, the carrier
density fluctuations in the two sheets would become less correlated.
%

%
\begin{figure}[!!!!!!!t!!!!!!!!!!!]
 \begin{center}
  \centering
   \includegraphics[width=8.5cm]{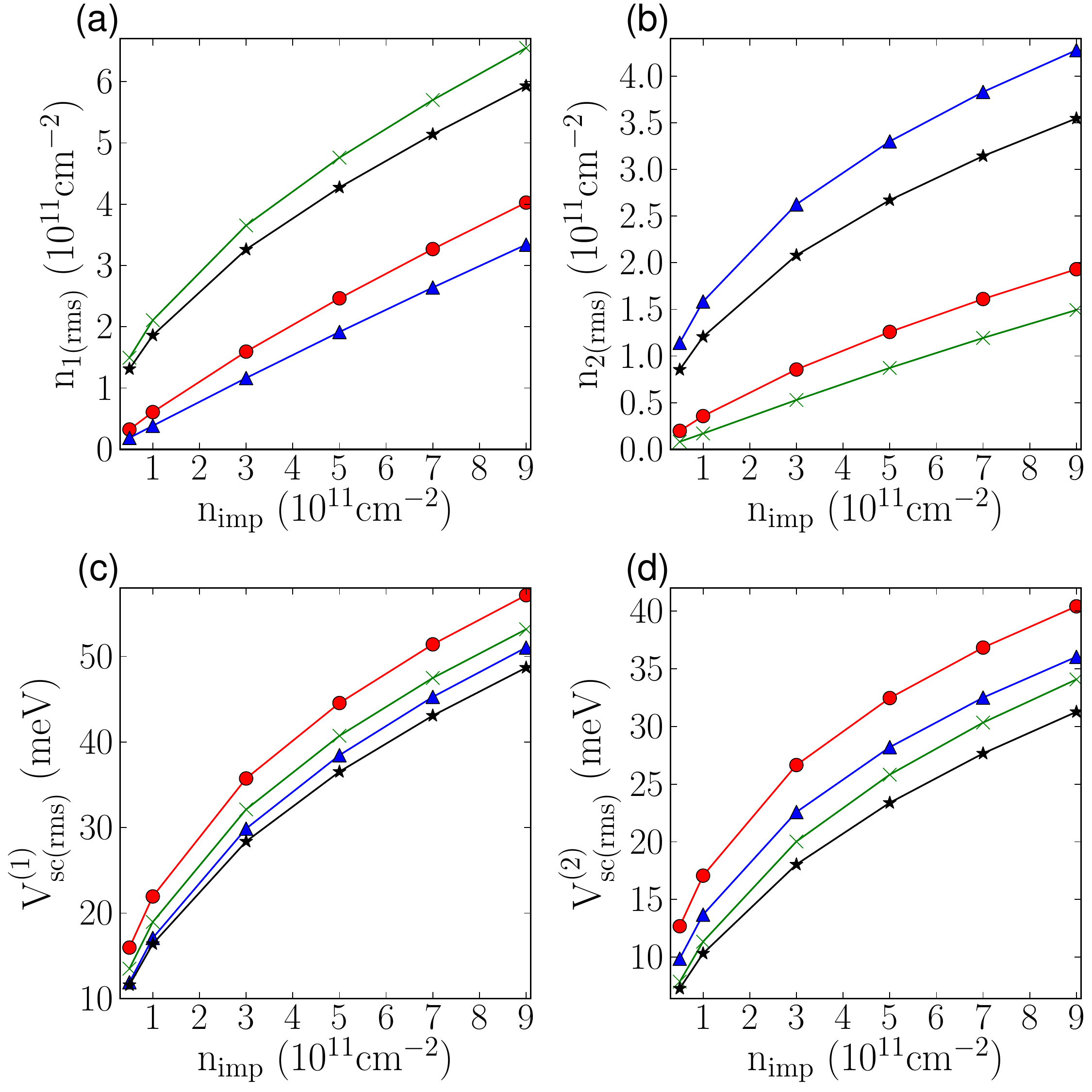}
   \caption{
           (Color online). Plots of (a) $\rm{n}_{1 \; (\rm{rms})}$, (b) $\rm{n}_{2 \; (\rm{rms})}$, (c) $\rm{V}^{(1)}_{sc \; (\rm{rms})}$, and (d) $\rm{V}^{(2)}_{sc \; (\rm{rms})}$ as a function of the impurity strength n$_{\rm{imp}}$ for the SLG-SLG system, d$ = 1 $ nm, d$_{12} = 1$ nm, and for four different carrier density averages. The circle symbols correspond to $\langle \rm{n}_1 \rangle = 0$ cm$^{-2}$ and $\langle \rm{n}_2 \rangle = 0$ cm$^{-2}$, the cross symbols to $\langle \rm{n}_1 \rangle = 5\times 10^{11}$ cm$^{-2}$ and $\langle \rm{n}_2 \rangle = 0$ cm$^{-2}$, the triangle symbols to $\langle \rm{n}_1 \rangle = 0$ cm$^{-2}$ and $\langle \rm{n}_2 \rangle = 5\times 10^{11}$ cm$^{-2}$, and the star symbols correspond to $\langle \rm{n}_1 \rangle =  5\times 10^{11}$ cm$^{-2}$ and $\langle \rm{n}_2 \rangle = 5\times 10^{11}$ cm$^{-2}$.
        } 
  \label{fig:figure_12}
 \end{center}
\end{figure}
\begin{figure}[!!!!t!!!!!!]
 \begin{center}
  \centering
   \includegraphics[width=8.5cm]{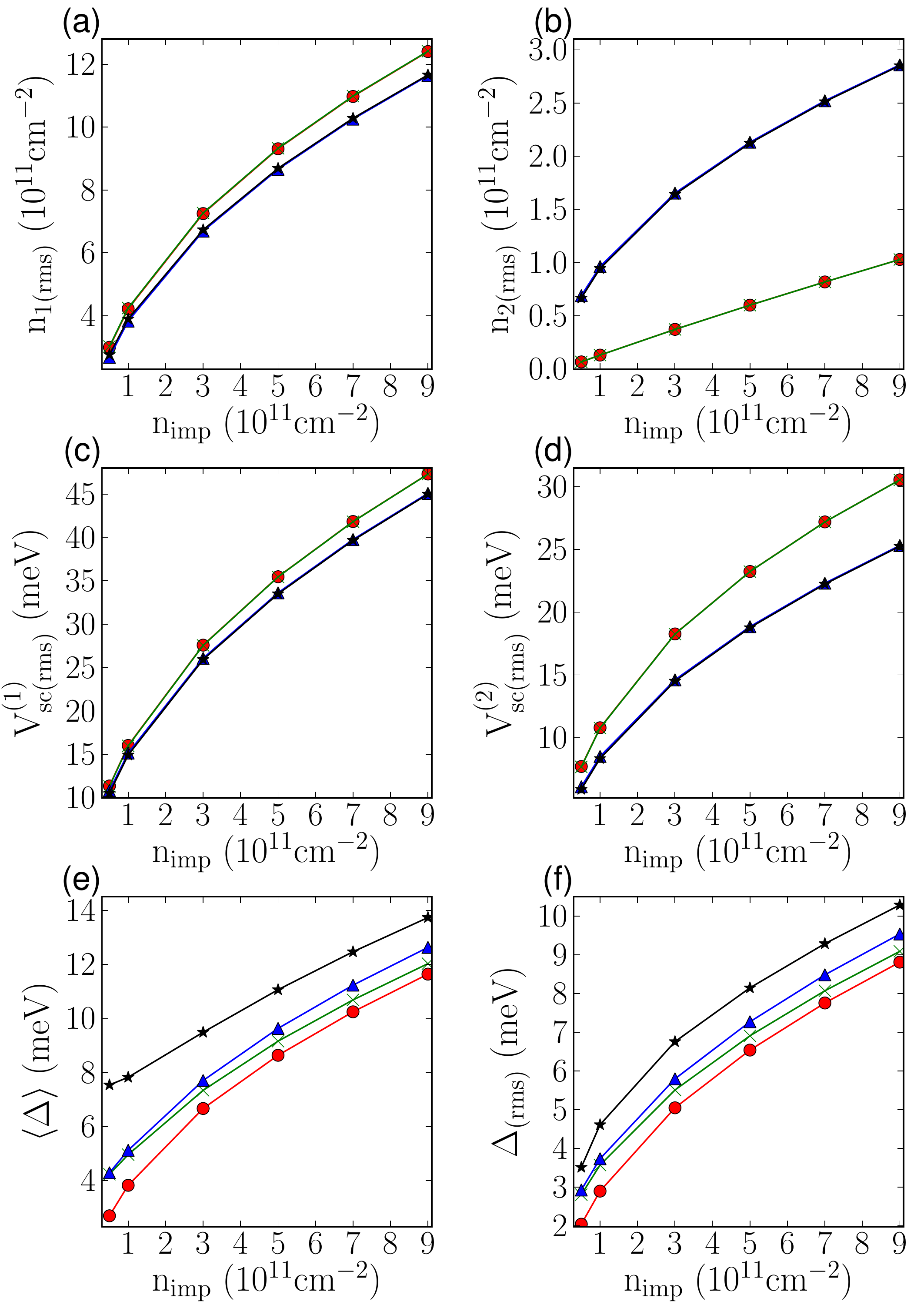}
   \caption{
           (Color online). Plots of (a) $\rm{n}_{1 \; (\rm{rms})}$, (b) $\rm{n}_{2 \; (\rm{rms})}$, (c) $\rm{V}^{(1)}_{sc \; (\rm{rms})}$, (d) $\rm{V}^{(2)}_{sc \; (\rm{rms})}$, (e) $\langle \Delta \rangle$, and (f) $\Delta_{(\rm{rms})}$ as a function of the impurity strength n$_{\rm{imp}}$ for the BLG-SLG system, d$ = 1 $ nm, d$_{12} = 1$ nm, and for four different carrier density averages.  The circle symbols correspond to $\langle \rm{n}_1 \rangle = 0$ cm$^{-2}$ and $\langle \rm{n}_2 \rangle = 0$ cm$^{-2}$, the cross symbols to $\langle \rm{n}_1 \rangle = 5\times 10^{11}$ cm$^{-2}$ and $\langle \rm{n}_2 \rangle = 0$ cm$^{-2}$, the triangle symbols to $\langle \rm{n}_1 \rangle = 0$ cm$^{-2}$ and $\langle \rm{n}_2 \rangle = 5\times 10^{11}$ cm$^{-2}$, and the star symbols correspond to $\langle \rm{n}_1 \rangle =  5\times 10^{11}$ cm$^{-2}$ and $\langle \rm{n}_2 \rangle = 5\times 10^{11}$ cm$^{-2}$.
        } 
  \label{fig:figure_13}
 \end{center}
\end{figure}

\begin{figure}[!!!!!!t!!!!!!]
 \begin{center}
  \centering
   \includegraphics[width=8.5cm]{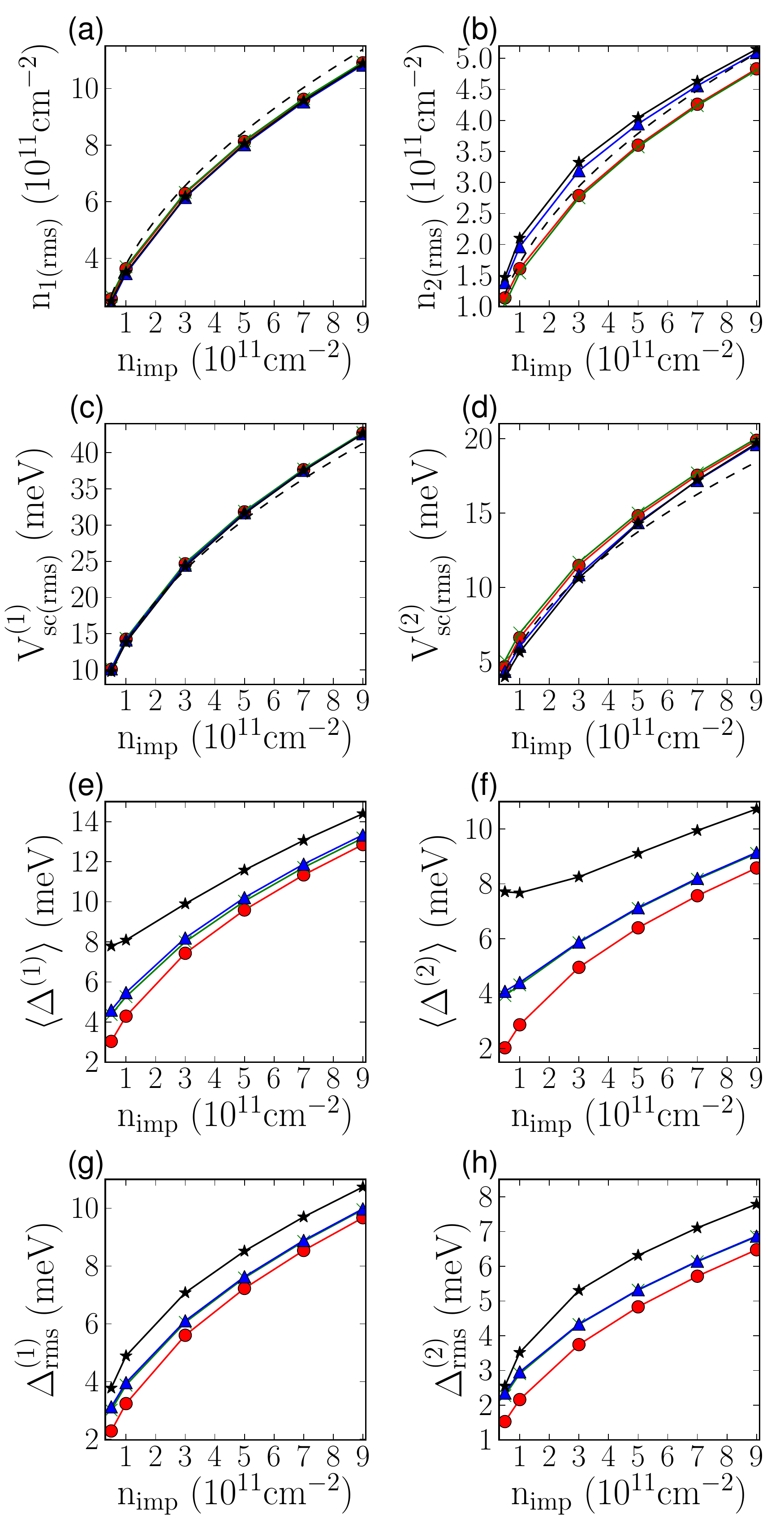}
   \caption{
           (Color online). Plots of (a) $\rm{n}_{1 \; (\rm{rms})}$, (b) $\rm{n}_{2 \; (\rm{rms})}$, (c) $\rm{V}^{(1)}_{sc \; (\rm{rms})}$, (d) $\rm{V}^{(2)}_{sc \; (\rm{rms})}$, (e) $\langle \Delta^{(1)} \rangle$, (f) $\langle \Delta^{(2)} \rangle$, (g) $\Delta^{(1)}_{(\rm{rms})}$, and $\Delta^{(2)}_{(\rm{rms})}$ as a function of the impurity strength n$_{\rm{imp}}$ for the BLG-BLG system, d$= 1 $ nm, d$_{12} = 1$ nm, and for four different carrier density averages. The circle symbols correspond to $\langle \rm{n}_1 \rangle = 0$ cm$^{-2}$ and $\langle \rm{n}_2 \rangle = 0$ cm$^{-2}$, the cross symbols to $\langle \rm{n}_1 \rangle = 5\times 10^{11}$ cm$^{-2}$ and $\langle \rm{n}_2 \rangle = 0$ cm$^{-2}$, the triangle symbols to $\langle \rm{n}_1 \rangle = 0$ cm$^{-2}$ and $\langle \rm{n}_2 \rangle = 5\times 10^{11}$ cm$^{-2}$, and the star symbols correspond to $\langle \rm{n}_1 \rangle =  5\times 10^{11}$ cm$^{-2}$ and $\langle \rm{n}_2 \rangle = 5\times 10^{11}$ cm$^{-2}$.
        } 
  \label{fig:figure_14}
 \end{center}
\end{figure}
%

Figures~\ref{fig:figure_12}-\ref{fig:figure_14} show the dependence on the impurity density 
of the statistical quantities characterizing the disordered the ground state,
for SLG-SLG, BLG-SLG, and BLG-BLG respectively.
To obtain these results we considered four different combination of average densities in
the two layers:
$(\nava,\navb)=(0,0); (5\times 10^{11}{\rm cm}^{-2}, 0), (0, 5\times 10^{11}{\rm cm}^{-2}, 0) (5\times 10^{11}{\rm cm}^{-2}, 5\times 10^{11}{\rm cm}^{-2})$.

For SLG-SLG, Fig.~\ref{fig:figure_12}, we have that the scaling with $\nimp$ is qualitatively similar for all four pairs of $(\nava,\navb)$ considered. 
The main feature is that, as is the case also for isolated SLG, $\nrms$ is lower for $\nav\approx 0$ than for 
$\nav$ away from the charge neutrality point. 
When the band structure of SLG is gapped we have that the scaling $\nrms$ and $\vrms$ with $\nimp$, Figs.~\ref{fig:figure_12_b}~(a)-(d), is qualitatively similar
to the one obtained for the gapless case.
For low values of $\nava$ ($\navb$) the fraction of the insulating area in layer 1 (2) depends quit strongly
on $\nimp$, as shown in Figs.~\ref{fig:figure_12_b}~(e),~(f). In addition we see that at low doping in layer 1 (2),
and low impurity densities, $\aia$ ($\aib$) depends quite strongly on $\nava$ ($\navb$), i.e. on the doping
of the other graphenic layer. 
For BLG-SLG heterostructures,
Fig.~\ref{fig:figure_13}, we find that $\nrms$ and $\vrms$ depend very weakly
on the $\nava$, consistent with the results shown in Fig.~\ref{fig:figure_8}.
The results of  Fig.~\ref{fig:figure_13}~(c) and (e) also show that
the ratio between the screened disorder potential and the average
band gap increases with $\nimp$. We therefore expect that the effects on the transport
properties due to the presence of a band gap 
\cite{oostinga2007,taychatanapat2010,zou2010,yan2010,rossi2011}
will be stronger for cleaner samples.
%

%
\begin{figure}[htb]
 \begin{center}
  \centering
   \includegraphics[width=8.5cm]{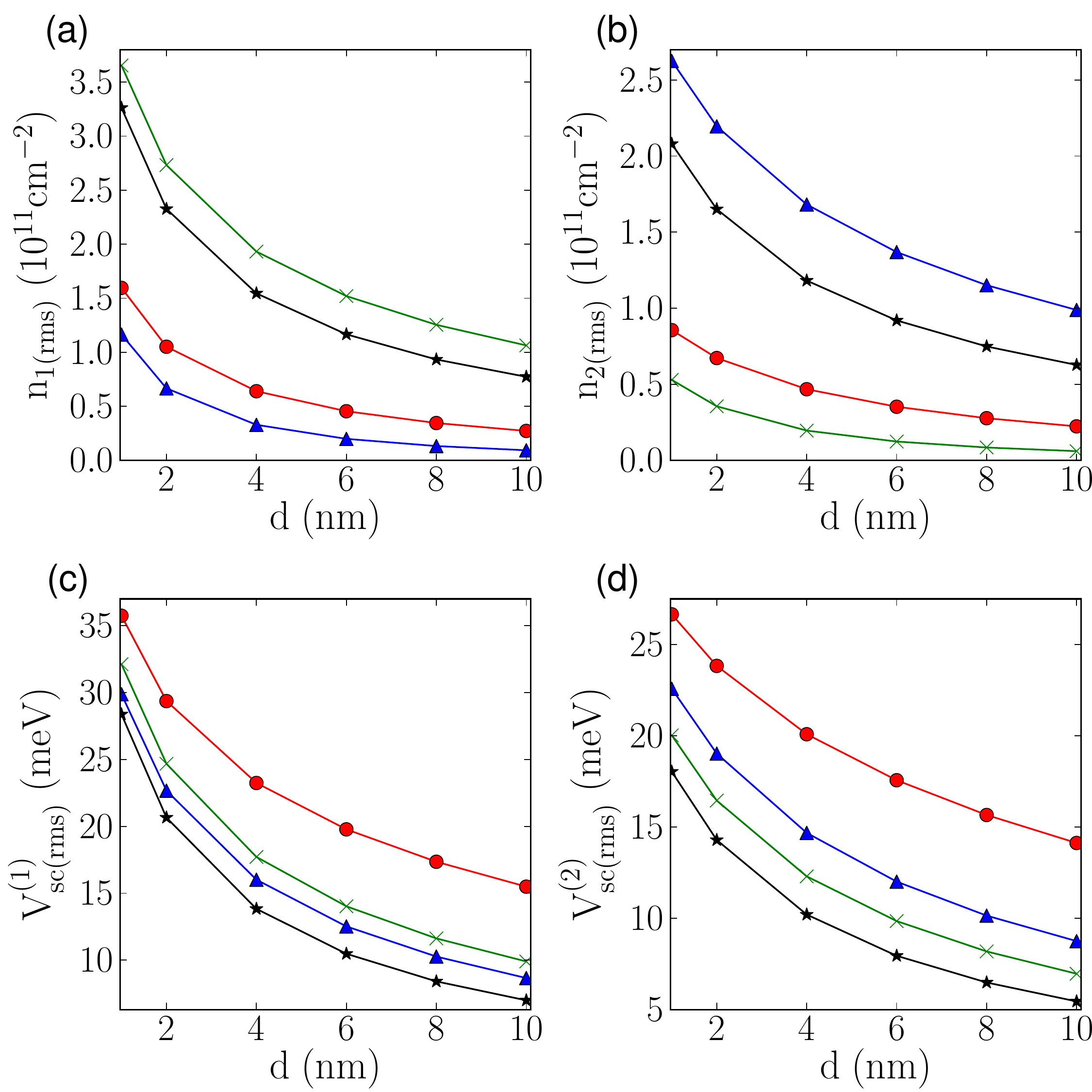}
   \caption{
           (Color online). Plots of (a) $\rm{n}_{1 \; (\rm{rms})}$, (b) $\rm{n}_{2 \; (\rm{rms})}$, (c) $\rm{V}^{(1)}_{sc \; (\rm{rms})}$, and (d) $\rm{V}^{(2)}_{sc \; (\rm{rms})}$ as a function of the distance between the impurities and the lower graphenic layer d for the SLG-SLG system, d$_{12} = 1 $ nm, and n$_{\rm{imp}} = 3 \times 10^{11}$ cm$^{-2}$. The circle symbols correspond to $\langle \rm{n}_1 \rangle = 0$ cm$^{-2}$ and $\langle \rm{n}_2 \rangle = 0$ cm$^{-2}$, the cross symbols to $\langle \rm{n}_1 \rangle = 5\times 10^{11}$ cm$^{-2}$ and $\langle \rm{n}_2 \rangle = 0$ cm$^{-2}$, the triangle symbols to $\langle \rm{n}_1 \rangle = 0$ cm$^{-2}$ and $\langle \rm{n}_2 \rangle = 5\times 10^{11}$ cm$^{-2}$, and the star symbols correspond to $\langle \rm{n}_1 \rangle =  5\times 10^{11}$ cm$^{-2}$ and $\langle \rm{n}_2 \rangle = 5\times 10^{11}$ cm$^{-2}$.
        } 
  \label{fig:figure_15}
 \end{center}
\end{figure}

Consistently with the results of Fig.~\ref{fig:figure_9} we find that 
for BLG-BLG systems the dependence of $\nrms$ and $\vrms$ on $\nimp$
is only weakly affected by the values 
of $\nava$ and $\navb$, Fig.~\ref{fig:figure_14}.
In Fig.~\ref{fig:figure_14}~(a)-(d) the dashed line shows the results obtained
equations 
\ceq{eq:nrms-gapless-blg}
\ceq{eq:vrms-gapless-blg}
obtained assuming $\Delta=0$. We see that, {\em for the purpose
of estimating $\nrms$ and $\vrms$}, in BLG-BLG heterostructures
neglecting the presence of a band-gap returns results that
are in good agreement with the results obtained taking
into account the fact that $\Delta\neq 0$.
As in BLG-SLG systems we observe that also in BLG-BLG heterostructures
the ratio $\vrms/\Deltaav$ increases with $\nimp$.
However, we notice that for the top BLG layer 
there is a large range of values of $\nimp$, and dopings, for which
$\Deltaav$ is larger than $\vrms$ and for which we therefore expect
the top layer to behave as an insulator.

\begin{figure}[htb]
 \begin{center}
  \centering
   \includegraphics[width=8.5cm]{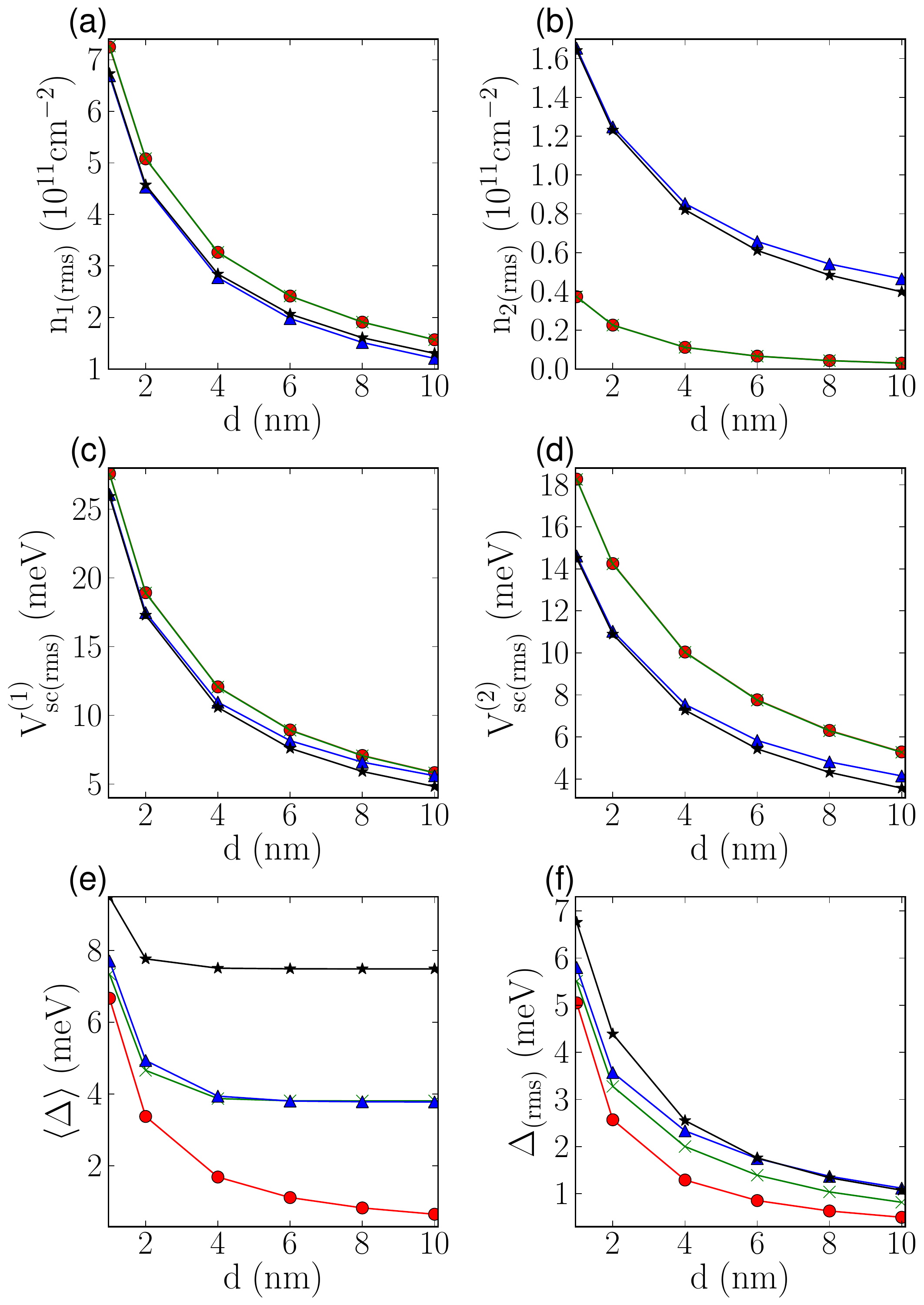}
   \caption{
           (Color online). Plots of (a) $\rm{n}_{1 \; (\rm{rms})}$, (b) $\rm{n}_{2 \; (\rm{rms})}$, (c) $\rm{V}^{(1)}_{sc \; (\rm{rms})}$, (d) $\rm{V}^{(2)}_{sc \; (\rm{rms})}$, (e) $\langle \Delta \rangle$, and (f) $\Delta_{(\rm{rms})}$ as a function of d for the BLG-SLG system, d$_{12} = 1 $ nm, n$_{\rm{imp}} = 3 \times 10^{11}$ cm$^{-2}$, and for four different carrier density averages.  The circle symbols correspond to $\langle \rm{n}_1 \rangle = 0$ cm$^{-2}$ and $\langle \rm{n}_2 \rangle = 0$ cm$^{-2}$, the cross symbols to $\langle \rm{n}_1 \rangle = 5\times 10^{11}$ cm$^{-2}$ and $\langle \rm{n}_2 \rangle = 0$ cm$^{-2}$, the triangle symbols to $\langle \rm{n}_1 \rangle = 0$ cm$^{-2}$ and $\langle \rm{n}_2 \rangle = 5\times 10^{11}$ cm$^{-2}$, and the star symbols correspond to $\langle \rm{n}_1 \rangle =  5\times 10^{11}$ cm$^{-2}$ and $\langle \rm{n}_2 \rangle = 5\times 10^{11}$ cm$^{-2}$.
        } 
  \label{fig:figure_16}
 \end{center}
\end{figure}
%

\begin{figure}[!!!!!!h!!!!!!tb]
 \begin{center}
  \centering
   \includegraphics[width=8.5cm]{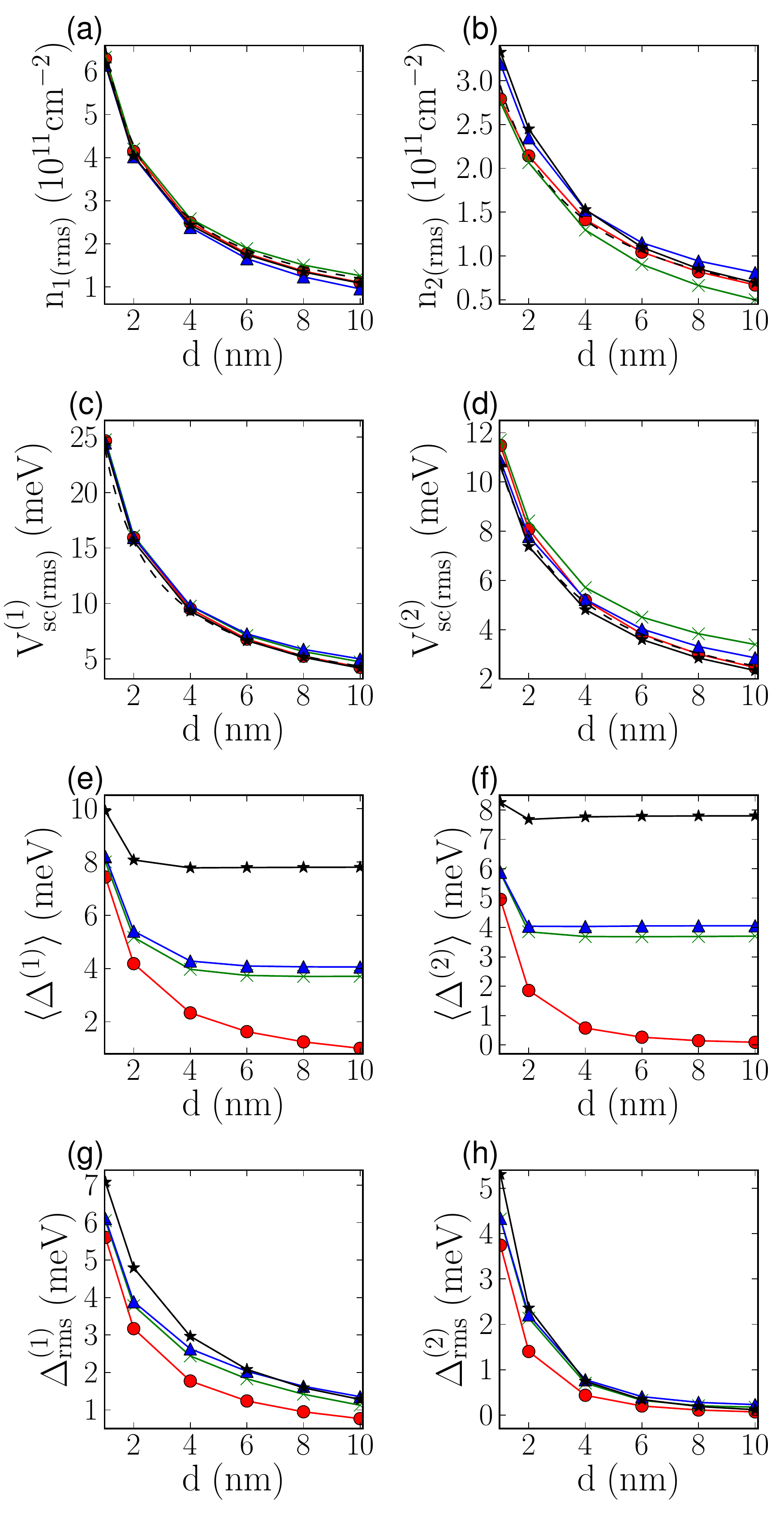}
   \caption{
           (Color online). Plots of (a) $\rm{n}_{1 \; (\rm{rms})}$, (b) $\rm{n}_{2 \; (\rm{rms})}$, (c) $\rm{V}^{(1)}_{sc \; (\rm{rms})}$, (d) $\rm{V}^{(2)}_{sc \; (\rm{rms})}$, (e) $\langle \Delta^{(1)} \rangle$, (f) $\langle \Delta^{(2)} \rangle$, (g) $\Delta^{(1)}_{(\rm{rms})}$, and $\Delta^{(2)}_{(\rm{rms})}$ as a function d for the BLG-BLG system, d$_{12}= 1 $ nm, n$_{\rm{imp}} = 3 \times 10^{11}$ cm$^{-2}$, and for four different carrier density averages. The circle symbols correspond to $\langle \rm{n}_1 \rangle = 0$ cm$^{-2}$ and $\langle \rm{n}_2 \rangle = 0$ cm$^{-2}$, the cross symbols to $\langle \rm{n}_1 \rangle = 5\times 10^{11}$ cm$^{-2}$ and $\langle \rm{n}_2 \rangle = 0$ cm$^{-2}$, the triangle symbols to $\langle \rm{n}_1 \rangle = 0$ cm$^{-2}$ and $\langle \rm{n}_2 \rangle = 5\times 10^{11}$ cm$^{-2}$, and the star symbols correspond to $\langle \rm{n}_1 \rangle =  5\times 10^{11}$ cm$^{-2}$ and $\langle \rm{n}_2 \rangle = 5\times 10^{11}$ cm$^{-2}$.
        } 
  \label{fig:figure_17}
 \end{center}
\end{figure}

As the distance $d$ of the charge impurities from the bottom layer is increased, 
the amplitude of the carrier density inhomogeneities and of the
r.m.s. of the screened disorder decrease rapidly for all the
three heterostructures considered. This is shown in 
Figs.~\ref{fig:figure_15}-\ref{fig:figure_17}.
In particular, panel (d) of these figures shows
that for $d\gtrsim 10$~nm, $\vrms$ in the top layer
is extremely small, smaller than 5~meV for the realistic
parameter considered. These results suggest
that the combination of first screening layer (graphenic or metallic)
and a clean buffer layer of a high quality dielectric,
such as hexagonal boron nitride (hBN),
10~nm thick or more would reduce the effects of
the disorder due to charge impurities to almost negligible levels.
%

For BLG-BLG systems we find that the scaling of $\nrms$ 
and $\vrms$ with $d$, analogously as for the scaling with $\nimp$,
is very well approximated by 
equations
\ceq{eq:nrms-gapless-blg},
\ceq{eq:vrms-gapless-blg}
derived in the limit $\Delta=0$.
Also, we find that for $d \gtrsim 3$~nm $\Deltaav$ dependence on $d$
is very weak, and that the ratio $\Deltarms/\Deltaav$ is quite small.
This is due to the fact that as $d$ increases the disorder potential
provides a decreasing contribution to the perpendicular electric field
and therefore to the band-gap of BLG. For very large $d$ and $\nava$ (and/or $\navb$) not zero the finite
value of the band-gap is due to the almost uniform charge distributions 
in the graphenic layers and metal gates.

\begin{figure}[htb]
 \begin{center}
  \centering
   \includegraphics[width=8.5cm]{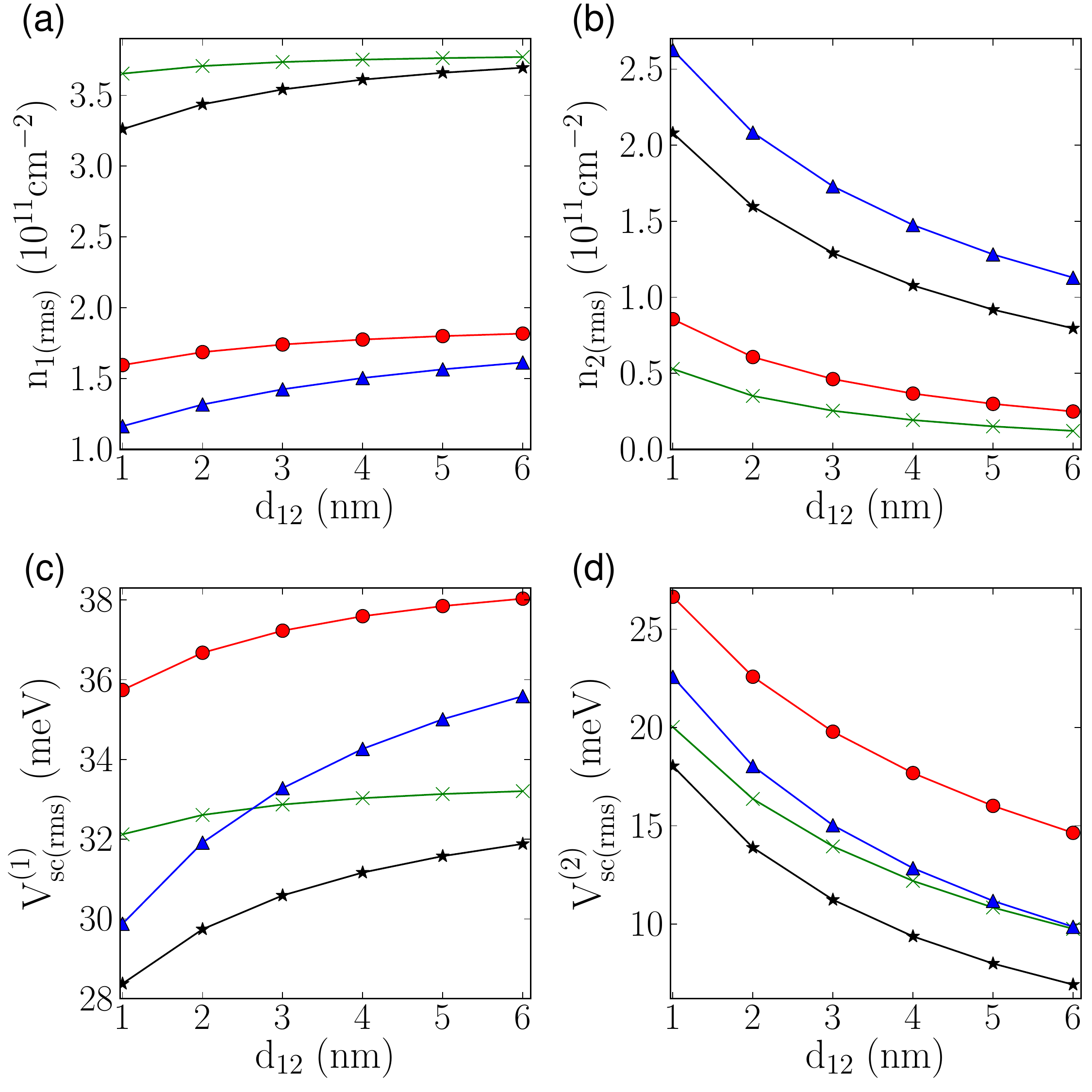}
   \caption{
           (Color online). Plots of (a) $\rm{n}_{1 \; (\rm{rms})}$, (b) $\rm{n}_{2 \; (\rm{rms})}$, (c) $\rm{V}^{(1)}_{sc \; (\rm{rms})}$, and (d) $\rm{V}^{(2)}_{sc \; (\rm{rms})}$ as a function of the distance between graphenic layers d$_{12}$ for the SLG-SLG system, d$ = 1 $ nm, and n$_{\rm{imp}} = 3 \times 10^{11}$ cm$^{-2}$. The circle symbols correspond to $\langle \rm{n}_1 \rangle = 0$ cm$^{-2}$ and $\langle \rm{n}_2 \rangle = 0$ cm$^{-2}$, the cross symbols to $\langle \rm{n}_1 \rangle = 5\times 10^{11}$ cm$^{-2}$ and $\langle \rm{n}_2 \rangle = 0$ cm$^{-2}$, the triangle symbols to $\langle \rm{n}_1 \rangle = 0$ cm$^{-2}$ and $\langle \rm{n}_2 \rangle = 5\times 10^{11}$ cm$^{-2}$, and the star symbols correspond to $\langle \rm{n}_1 \rangle =  5\times 10^{11}$ cm$^{-2}$ and $\langle \rm{n}_2 \rangle = 5\times 10^{11}$ cm$^{-2}$.
        } 
  \label{fig:figure_18}
 \end{center}
\end{figure}
\begin{figure}[htb]
 \begin{center}
  \centering
   \includegraphics[width=8.5cm]{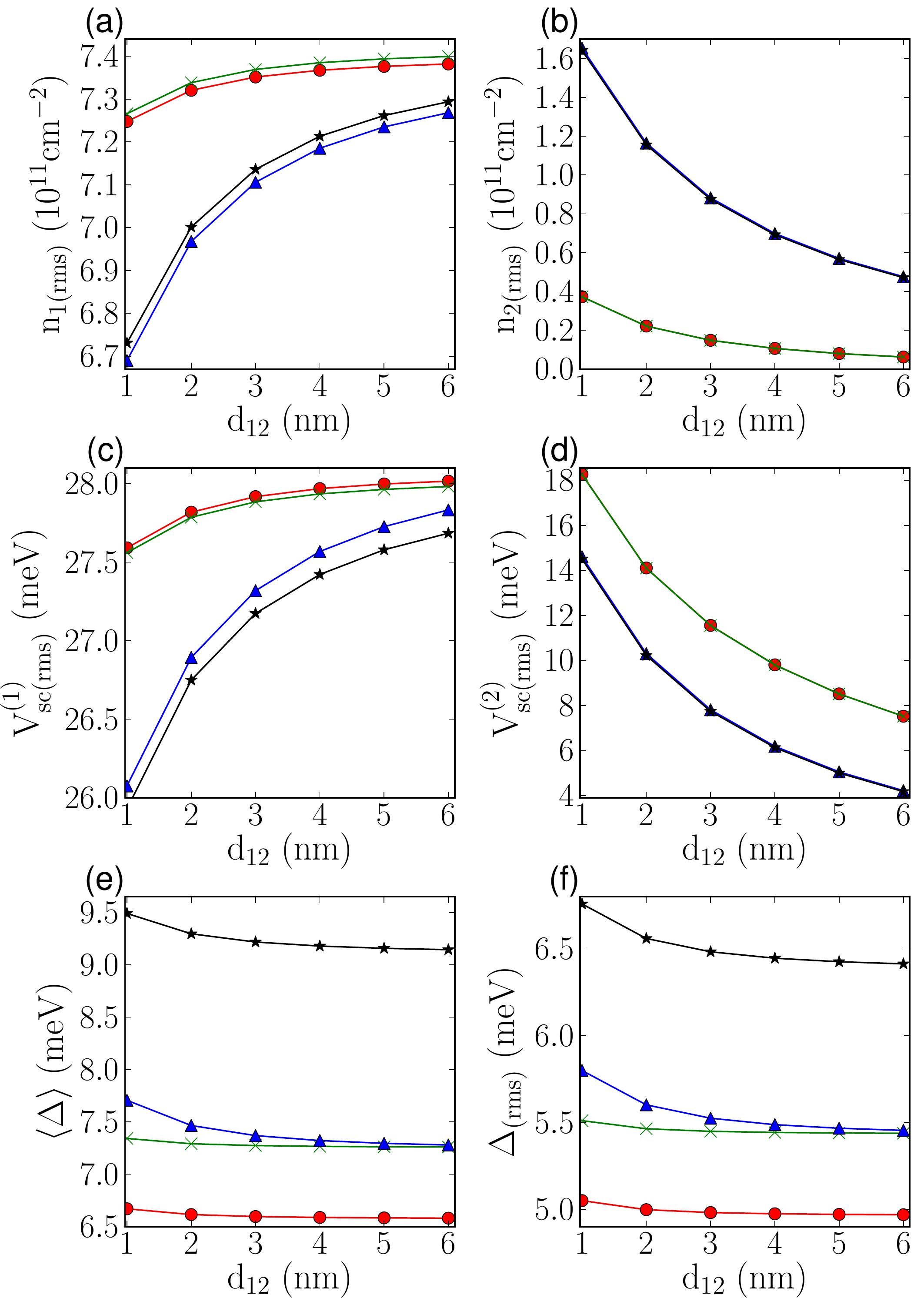}
   \caption{
           (Color online). Plots of (a) $\rm{n}_{1 \; (\rm{rms})}$, (b) $\rm{n}_{2 \; (\rm{rms})}$, (c) $\rm{V}^{(1)}_{sc \; (\rm{rms})}$, (d) $\rm{V}^{(2)}_{sc \; (\rm{rms})}$, (e) $\langle \Delta \rangle$, and (f) $\Delta_{(\rm{rms})}$ as a function of d$_{12}$ for the BLG-SLG system, d$ = 1 $ nm, n$_{\rm{imp}} = 3 \times 10^{11}$ cm$^{-2}$, and for four different carrier density averages.  The circle symbols correspond to $\langle \rm{n}_1 \rangle = 0$ cm$^{-2}$ and $\langle \rm{n}_2 \rangle = 0$ cm$^{-2}$, the cross symbols to $\langle \rm{n}_1 \rangle = 5\times 10^{11}$ cm$^{-2}$ and $\langle \rm{n}_2 \rangle = 0$ cm$^{-2}$, the triangle symbols to $\langle \rm{n}_1 \rangle = 0$ cm$^{-2}$ and $\langle \rm{n}_2 \rangle = 5\times 10^{11}$ cm$^{-2}$, and the star symbols correspond to $\langle \rm{n}_1 \rangle =  5\times 10^{11}$ cm$^{-2}$ and $\langle \rm{n}_2 \rangle = 5\times 10^{11}$ cm$^{-2}$.
        } 
  \label{fig:figure_19}
 \end{center}
\end{figure}
%

\begin{figure}[!!!!!t]
 \begin{center}
  \centering
   \includegraphics[width=8.5cm]{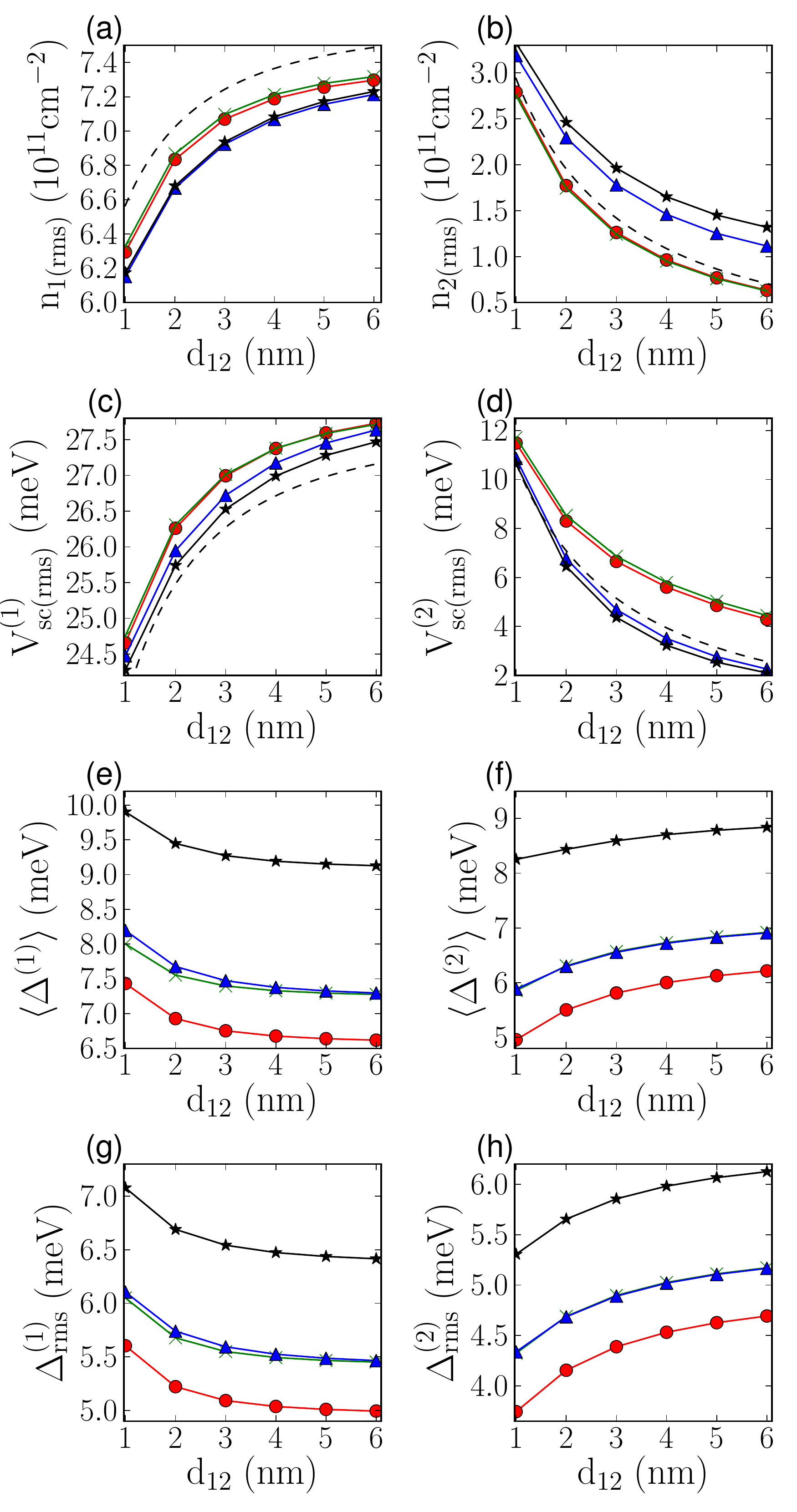}
   \caption{
           (Color online). Plots of (a) $\rm{n}_{1 \; (\rm{rms})}$, (b) $\rm{n}_{2 \; (\rm{rms})}$, (c) $\rm{V}^{(1)}_{sc \; (\rm{rms})}$, (d) $\rm{V}^{(2)}_{sc \; (\rm{rms})}$, (e) $\langle \Delta^{(1)} \rangle$, (f) $\langle \Delta^{(2)} \rangle$, (g) $\Delta^{(1)}_{(\rm{rms})}$, and $\Delta^{(2)}_{(\rm{rms})}$ as a function d$_{12}$ for the BLG-BLG system, d$= 1 $ nm, n$_{\rm{imp}} = 3 \times 10^{11}$ cm$^{-2}$, and for four different carrier density averages. The circle symbols correspond to $\langle \rm{n}_1 \rangle = 0$ cm$^{-2}$ and $\langle \rm{n}_2 \rangle = 0$ cm$^{-2}$, the cross symbols to $\langle \rm{n}_1 \rangle = 5\times 10^{11}$ cm$^{-2}$ and $\langle \rm{n}_2 \rangle = 0$ cm$^{-2}$, the triangle symbols to $\langle \rm{n}_1 \rangle = 0$ cm$^{-2}$ and $\langle \rm{n}_2 \rangle = 5\times 10^{11}$ cm$^{-2}$, and the star symbols correspond to $\langle \rm{n}_1 \rangle =  5\times 10^{11}$ cm$^{-2}$ and $\langle \rm{n}_2 \rangle = 5\times 10^{11}$ cm$^{-2}$.
        } 
  \label{fig:figure_20}
 \end{center}
\end{figure}

Figures~\ref{fig:figure_18}-\ref{fig:figure_20} show the dependence
of $\nrms$, $\vrms$ and $\Delta$ on the distance, $d_{12}$, between
the two layers forming the heterostructure. 
For the SLG-SLG heterostructure, Fig.~\ref{fig:figure_18}, 
the scaling on $d_{12}$ of $\nrms$ and $\vrms$ in layer 1 (layer 2) depends strongly
on the average carrier density in layer 2 (layer 1). 
This is due to the fact that the ability of layer 1 (layer 2) to screen layer 2 (layer 1)
from the disorder potential depends strongly on its average carrier density.
For example, when $\navb=0$ layer 2 does not provide a significant
contribution to the screening of the disorder potential in layer 1
and therefore moving it away from layer 1, i.e. increasing $d_{12}$,
has only a very minor effect on the value of $\nrmsa$ and $\vrmsa$,
as shown in Fig.~\ref{fig:figure_18}~(a),~(b) respectively.

For BLG-SLG heterostructures, Fig.~\ref{fig:figure_19}, the dependence of $\nrms$ and $\vrms$ 
on $d_{12}$ it is almost independent of the average density in BLG, layer 1, a fact
that is consistent with the other results that we have presented above for BLG-SLG systems.
This reflects the fact that the density of states in BLG at low dopings depends only very weakly on the value of $\nav$.
As $d_{12}$ increases, the values of $\nrmsa$ and $\vrmsa$
approach asymptotically the values for isolated BLG.
Moreover, we observe that, as $d_{12}$ increases, the value of $\Deltaav$ and $\Deltarms$ approach a constant value,
independent of $d_{12}$, but dependent on $\navb$, Figs.~\ref{fig:figure_19}~(e),~(f).

%
%
This is due to the fact that as $d_{12}$ increases the screening effects of the top layer on the bottom
layer decrease, as mentioned above, and the perpendicular electric field reaches a value
that is almost independent of $d_{12}$, but still dependent on $\navb$.
In these conditions, $\Delta$ in layer 1 depends on layer 2 only via $\navb$.Also,
as $d_{12}$ increases, $\Deltarms$ in layer 1 approaches a constant value
corresponding to the value of $\Deltarms$ for an isolated BLG sheet with
average band-gap $\Deltaav$. 

The effect of a change in $d_{12}$ in BLG-BLG systems is shown in Fig.~\ref{fig:figure_20}.
In figures~\ref{fig:figure_20}~(a)-(d) the dashed lines show the results obtained using equations
\ceq{eq:nrms-gapless-blg},
\ceq{eq:vrms-gapless-blg}
obtained by setting $\Delta=0$ in both layers.
We see that for the dependence of $\nrms$ and $\vrms$ on $d_{12}$, as
for the dependence on $\nimp$ and $d$,
the results obtained by setting $\Delta=0$ are in good quantitative
agreement with the results obtained by calculating $\Delta$ self-consistently.
For the same reason mentioned for the case of BLG-SLG heterostructure, we find
that $\Deltaav$ and $\Deltarms$ in the bottom layer decrease with $d_{12}$ 
and approach a constant value for large $d_{12}$. 
As for BLG-SLG we see that as $d_{12}$ increases $\Deltarms$ takes
values that very close to the values of $\Deltaav$.
%
%
\begin{figure}[!ht]
 \begin{center}
  \centering
   \includegraphics[width=8.5cm]{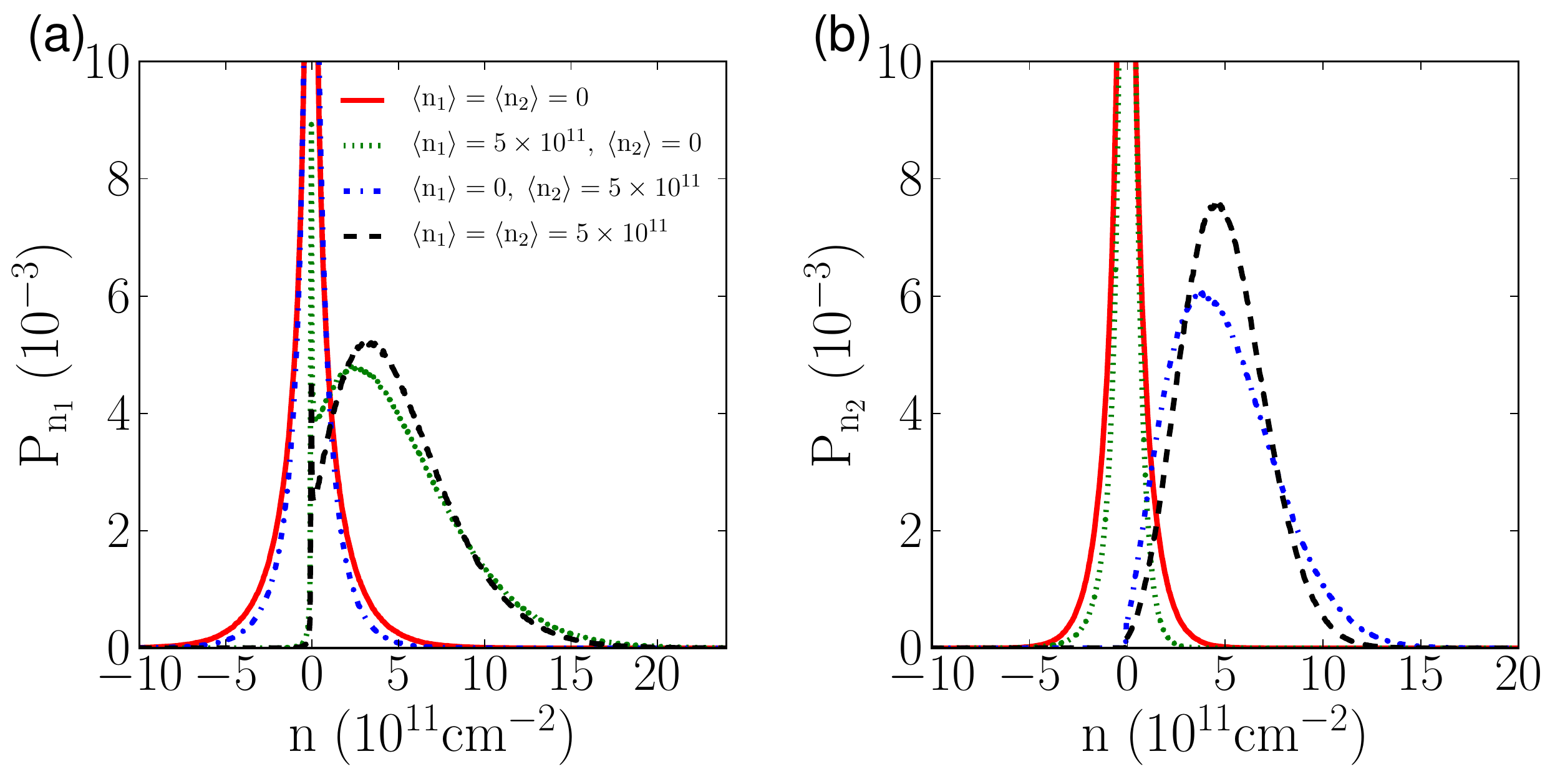}
   \caption{
           (Color online). Plots of the carrier density probability distribution (a) $P_{ \rm{n}_{1} }$, and (b) $P_{ \rm{n}_{2} }$, for the SLG-SLG system, d$ = 1 $ nm, and n$_{\rm{imp}} = 3 \times 10^{11}$ cm$^{-2}$. The solid line corresponds to $\langle \rm{n}_1 \rangle = 0$ cm$^{-2}$ and $\langle \rm{n}_2 \rangle = 0$ cm$^{-2}$, the dotted line corresponds to $\langle \rm{n}_1 \rangle = 5\times 10^{11}$ cm$^{-2}$ and $\langle \rm{n}_2 \rangle = 0$ cm$^{-2}$, the line-dotted curve corresponds to $\langle \rm{n}_1 \rangle = 0$ cm$^{-2}$ and $\langle \rm{n}_2 \rangle = 5\times 10^{11}$ cm$^{-2}$, and the dashed line corresponds to $\langle \rm{n}_1 \rangle =  5\times 10^{11}$ cm$^{-2}$ and $\langle \rm{n}_2 \rangle = 5\times 10^{11}$ cm$^{-2}$.
        } 
  \label{fig:figure_21}
 \end{center}
\end{figure}
\begin{figure}[!ht]
 \begin{center}
  \centering
   \includegraphics[width=8.5cm]{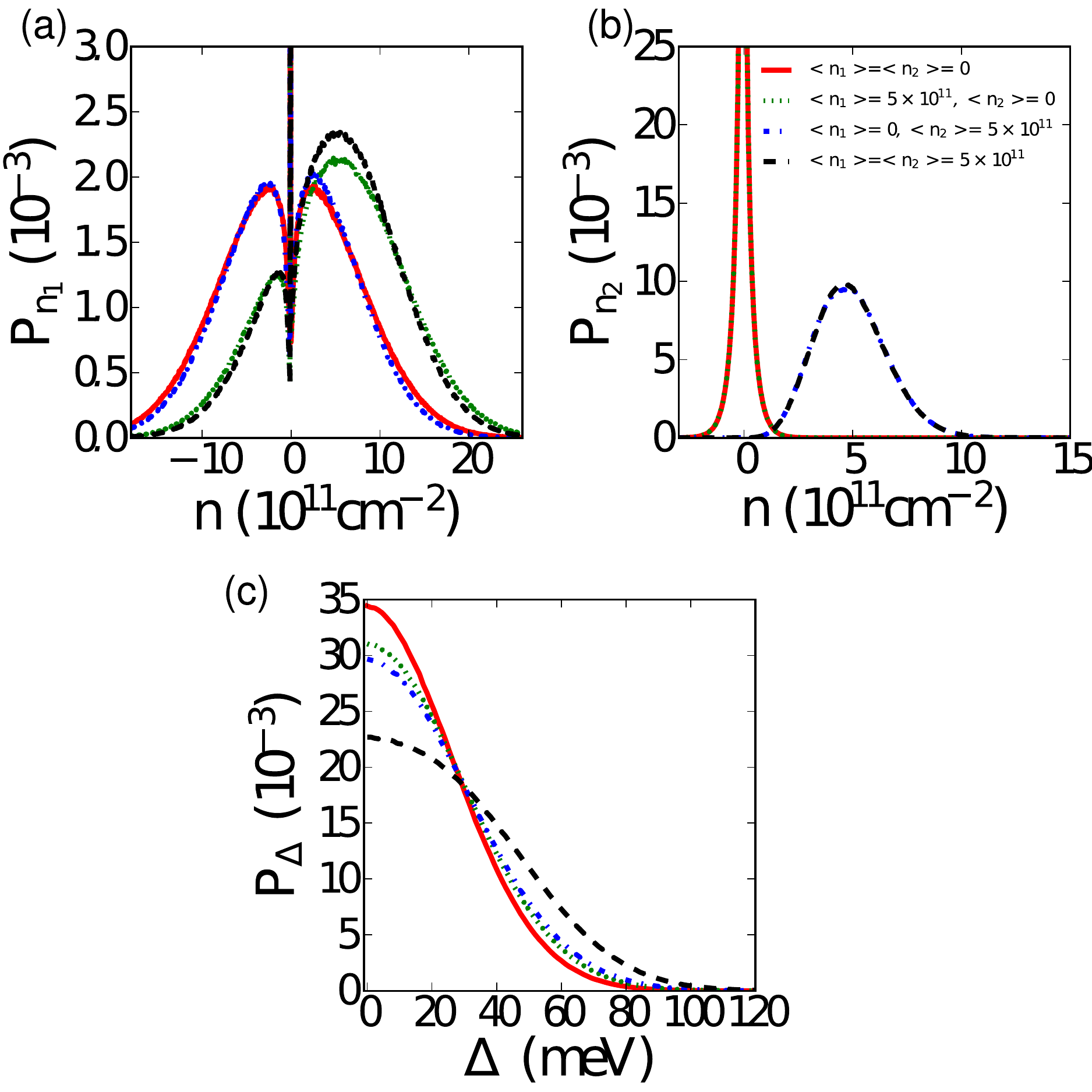}
   \caption{
           (Color online). Plots of the carrier density probability distribution (a) $P_{ \rm{n}_{1} }$, and (b) $P_{ \rm{n}_{2} }$, and plot of the gap probability distribution (c) $P_{ \rm{\Delta}}$ for the BLG-SLG system, d$ = 1 $ nm, and n$_{\rm{imp}} = 3 \times 10^{11}$ cm$^{-2}$. The solid line corresponds to $\langle \rm{n}_1 \rangle = 0$ cm$^{-2}$ and $\langle \rm{n}_2 \rangle = 0$ cm$^{-2}$, the dotted line corresponds to $\langle \rm{n}_1 \rangle = 5\times 10^{11}$ cm$^{-2}$ and $\langle \rm{n}_2 \rangle = 0$ cm$^{-2}$, the line-dotted curve corresponds to $\langle \rm{n}_1 \rangle = 0$ cm$^{-2}$ and $\langle \rm{n}_2 \rangle = 5\times 10^{11}$ cm$^{-2}$, and the dashed line corresponds to $\langle \rm{n}_1 \rangle =  5\times 10^{11}$ cm$^{-2}$ and $\langle \rm{n}_2 \rangle = 5\times 10^{11}$ cm$^{-2}$.
        } 
  \label{fig:figure_22}
 \end{center}
\end{figure}
\begin{figure}[!ht]
 \begin{center}
  \centering
   \includegraphics[width=8.5cm]{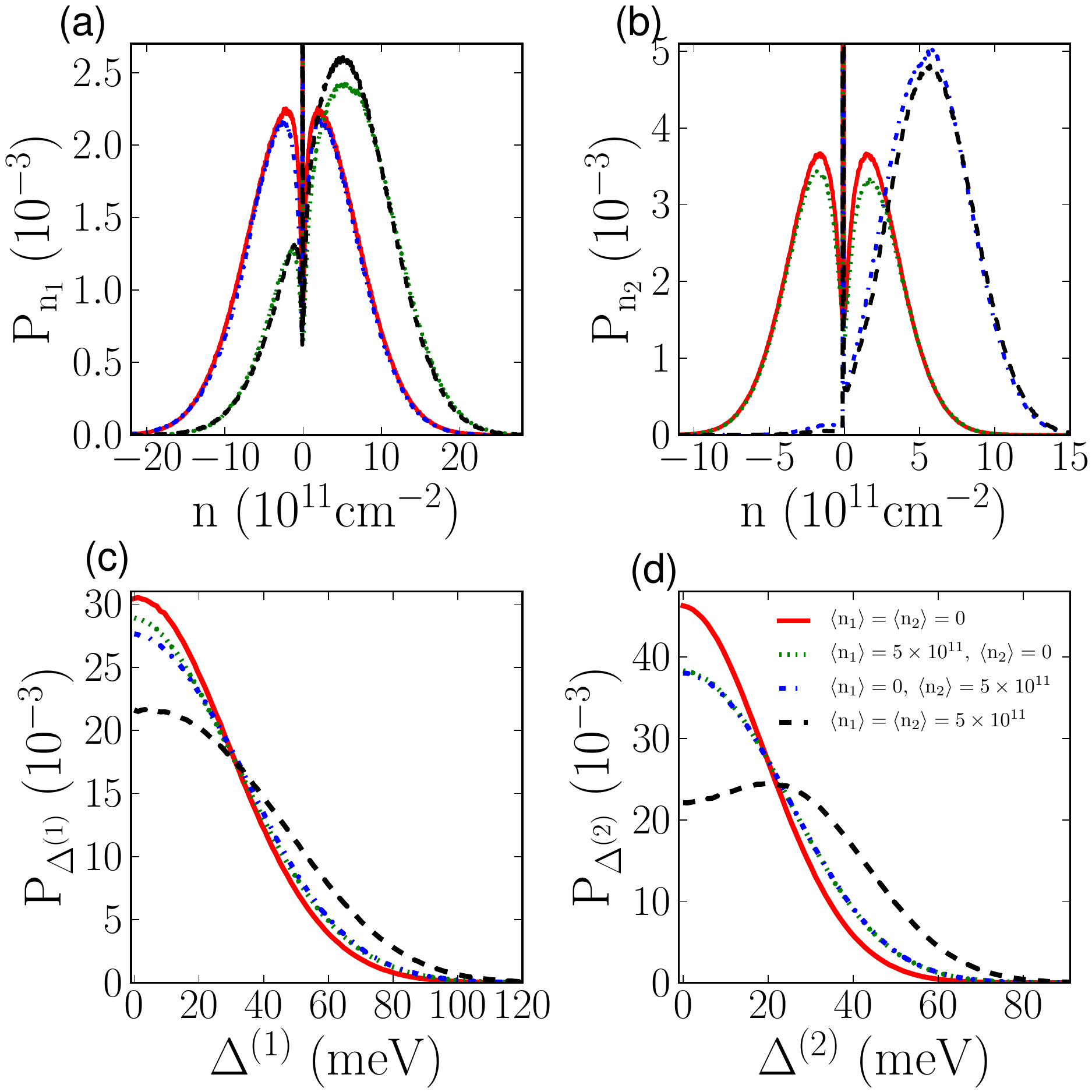}
   \caption{
           (Color online). Plots of the carrier density probability distribution (a) $P_{ \rm{n}_{1} }$, and (b) $P_{ \rm{n}_{2} }$, and plots of the gap probability distributions (c) $P_{ \rm{\Delta^{(1)}}}$, and (d) $P_{ \rm{\Delta^{(2)}}}$ for the BLG-BLG system, d$ = 1 $ nm, and n$_{\rm{imp}} = 3 \times 10^{11}$ cm$^{-2}$. The solid line corresponds to $\langle \rm{n}_1 \rangle = 0$ cm$^{-2}$ and $\langle \rm{n}_2 \rangle = 0$ cm$^{-2}$, the dotted line corresponds to $\langle \rm{n}_1 \rangle = 5\times 10^{11}$ cm$^{-2}$ and $\langle \rm{n}_2 \rangle = 0$ cm$^{-2}$, the line-dotted curve corresponds to $\langle \rm{n}_1 \rangle = 0$ cm$^{-2}$ and $\langle \rm{n}_2 \rangle = 5\times 10^{11}$ cm$^{-2}$, and the dashed line corresponds to $\langle \rm{n}_1 \rangle =  5\times 10^{11}$ cm$^{-2}$ and $\langle \rm{n}_2 \rangle = 5\times 10^{11}$ cm$^{-2}$.
        } 
  \label{fig:figure_23}
 \end{center}
\end{figure}

In figure~\ref{fig:figure_21} we show the probability distribution ($P_{n_i}$) 
for the carrier density in the two layers of a SLG-SLG heterostructure for different
values of the average doping $\nava$ and $\navb$.
For $\nava=0$ ($\navb=0$) we see that $P_{n_1}$ ($P_{n_2}$) is very strongly peaked
around the charge neutrality point: for $n_i\to 0$ $P_{n_1}$ reaches values that
are orders of magnitude outside the scale of the figures. In this situation
$P_{n_i}$ is not Gaussian.
As $\nava$ ($\navb$) increases $P_{n_1}$ ($P_{n_2}$) becomes bimodal: it exhibits a very strong and narrow
peak at $n_1=0$ ($n_2=0$) and a much broader peak around $n_1=\nava$ ($n_2=\navb)$.
Only for quite large values of $\nav$ $P_n$ is well approximated by a simple Gaussian
centered around $\nav$. 
The properties of $P_n$ for SLG-SLG heterostructures, and its dependence on $\nav$,
are very similar to the ones of an isolated layer of graphene \cite{rossi2008}. The only difference
is that, for the same strength of the disorder, the peaks of $P_n$ in the second
layer are narrower than in the first layer and than in an isolated graphene layer,
because of the screening of the disorder by the first layer. In addition we find that
because of the screening effect of the first layer, the value of $\navb$ above
which $P_{n_2}$ has a simple Gaussian peak centered around $\navb$ is lower
than for the first layer (and than for isolated graphene).

Figure~\ref{fig:figure_22}~(a),~(b) show the results for $P_{n_i}$ for the 
case of BLG-SLG. The presence of a perpendicular electric field induces the opening
of a band-gap in BLG. This causes the presence of small gapped regions with zero
carrier density. 
As a consequence $P_{n_1}$ exhibits an extremely narrow peak for $n_1=0$
surrounded by two large shoulders, Fig.~\ref{fig:figure_22}~(a).
As $\nava$ increases the narrow peak at $n_1=0$ decreases and
the two-shoulders structure becomes asymmetric evolving toward
a single, broad, Gaussian peak centered around $\nava$.
$P_{n}$ in the top layer, the SLG layer, is qualitatively very similar
to the $P_n$ of the top layer in SLG-SLG structures, just much narrower 
due to the fact that the BLG, as a bottom layer, is much more
efficient to screen the disorder potential.

Figure~\ref{fig:figure_22}~(c) shows the profile of the probability
distribution ($P_\Delta$) of the band gap in BLG.
We see that $P_\Delta$ has a Gaussian-like shape, approximately centered at zero
(of course limited to positive values). For the values
of $\nava$ and $\navb$ considered in Fig.~\ref{fig:figure_22}~(c)  
the profiles of $P_\Delta$ are qualitatively very similar indicating
that, for the cases shown, the main contribution to $\Delta$ is
due to the disorder potential. Only the profile for 
$\nava=\navb=5\times 10^{11}~{\rm cm}^{-2}$
shows a significant difference from the profiles for
the other cases. This is due to the fact that for $\nava=\navb=5\times 10^{11}~{\rm cm}^{-2}$
a uniform $\Delta$, independent of the disorder, is present that causes
a shift of the average value of $P_\Delta$.

Figures~\ref{fig:figure_23}~(a),~(b) show the results for $P_{n_i}$ for the 
case of BLG-BLG. The results are qualitatively similar to the results
shown in  Fig.~\ref{fig:figure_22}~(a) for the BLG layer of a BLG-SLG
structure, and the explanation of the main qualitative features
of $P_{n}$ presented for that case apply also here.
Figures~\ref{fig:figure_23}~(c),~(d) show $P_{\Delta}$ in the bottom
and top layer respectively.
In this case, for $\nava=\navb=5\times 10^{11}~{\rm cm}^{-2}$, especially
for the top layer, (black dashed line in Fig.~\ref{fig:figure_22}~(d),
it is clear that the average of $P_{\Delta}$ is shifted to the right
due to the fact that when $\nava\neq 0$ and/or $\navb\neq 0$
a uniform band-gap is present.

\section{On the metal-insulator transition in double-layer graphene heterostructures}
\label{sec:mit}

The experiments of Ref.~\onlinecite{ponomarenko2011} have shown that in SLG-SLG structures a density-tuned metal-insulator transition (MIT) can be induced
in one of the SLG layers by tuning the doping in the other layer. The fact that the MIT in one layer is tuned by the doping in the other
layer strongly suggests that long-range disorder, and in particular the electron-hole puddles that such disorder induces,
play a dominant role in the physics of the MIT in SLG-SLG systems. 

In Ref.~\onlinecite{ponomarenko2011} it was proposed that the insulating behavior of a graphene layer in a SLG-SLG heterostructure
is due to strong Anderson localization made possible in the system perhaps due to strong inter-valley scattering. 
The ``control'' graphene layer provides additional screening of the disorder induced by charge impurities and therefore a reduction
of the  amplitude of the electron-hole puddles in the studied layer. In the scenario proposed in 
Ref.~\onlinecite{ponomarenko2011} the increase of the doping in the control layer can reduce
the strength of the carrier density inhomogeneities in the studied layer, increasing the resistivity \cite{dassarma2011} to allow the manifestation
of the strong Anderson localization.
In Ref.~\onlinecite{kechedzi2012} the tunability of localization effects in the studied layer via the doping of
the control layer is attributed to the dependence of the scattering rate due to charge impurities and the dephasing time in the studied layer on the doping in the control layer.


Ref.~\onlinecite{dassarma2012} proposed a completely different scenario to interpret the results of Ref.~\onlinecite{ponomarenko2011}.
In this scenario the dramatic increase of the resistivity, close to the CNP, in the studied layer, as a function of doping 
in the control layer is not due to Anderson localization, but to the fact that, as the amplitude of the
disorder-induced electron-hole puddles decreases, the resistivity at the CNP diverges since in SLG the density of states vanishes at the CNP.
One of the key observations of Ref.~\onlinecite{dassarma2012} is that, contrary to 
metals, in systems like graphene, at low dopings, higher mobility samples exhibit higher resistivity.
This agrees with the experimental results of Ref.~\onlinecite{ponomarenko2011} that show that of the two graphene layers
forming the heterostructure, the one with the higher mobility is the one exhibiting the highest resistivity at low dopings.

\begin{figure}[!htb]
 \begin{center}
  \centering
   \includegraphics[width=8.5cm]{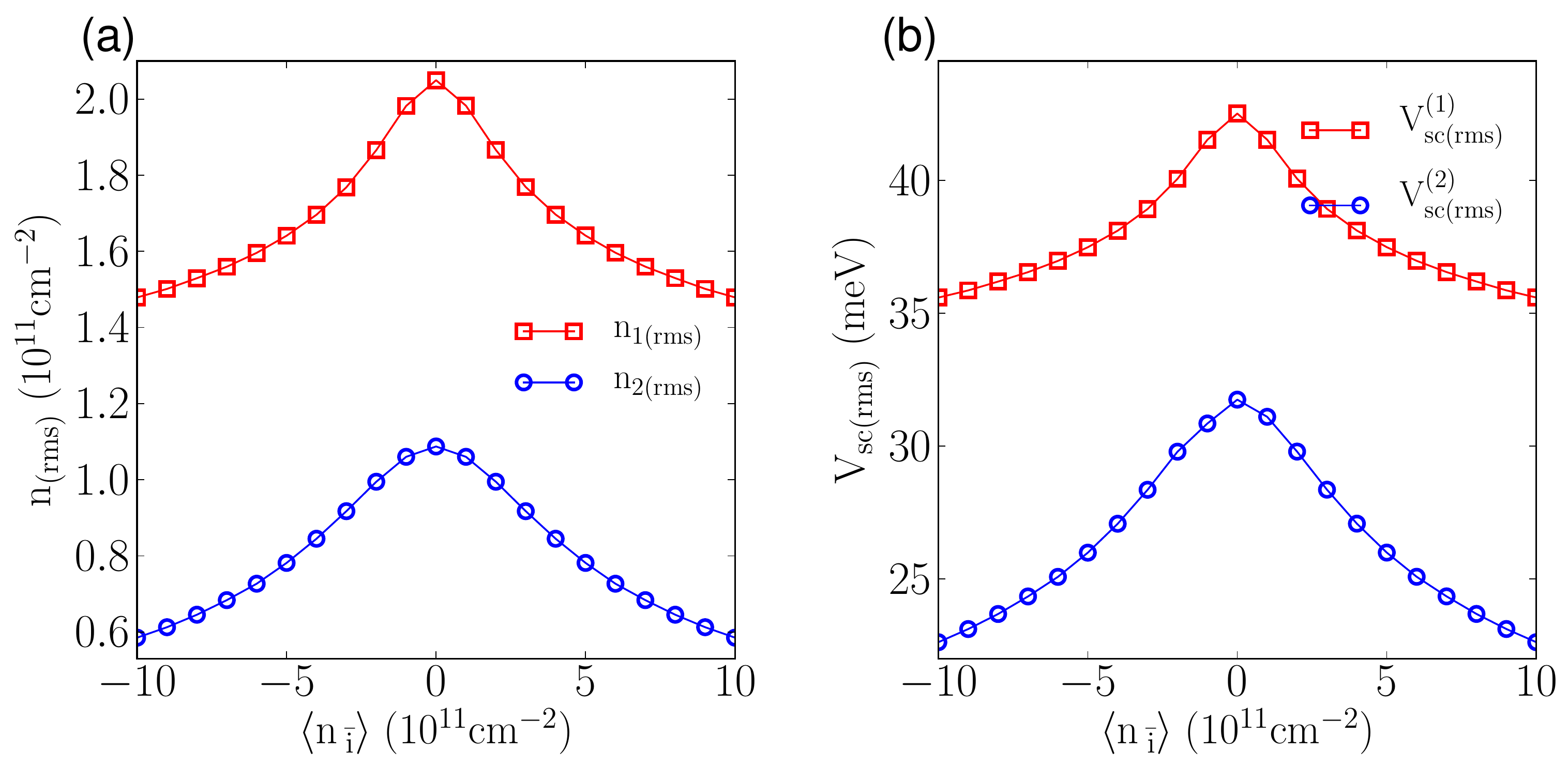}
   \caption{
           (Color online). Plots of (a) $\nrms$ and, (b) $\vrms$ at the CN in layer ``i'' as a function of the doping in the other layer $\langle \rm{n}_{\bar {\rm i}}  \rangle$, for d$ = 1$nm, d$_{12} = 1$nm and $\nimp= 3 \times 10^{11}$ cm$^{-2}$, for the gapped SLG-SLG heterostructure. The graphene spectrum has a gap equal to 20~meV. The squares correspond to the bottom SLG layer and the circles correspond to the top SLG layer. 
        } 
  \label{fig:figure_32g}
 \end{center}
\end{figure}

We note that the contrasting interpretations offered in Ref.~\onlinecite{ponomarenko2011} and Ref.~\onlinecite{dassarma2012} for the experimental observations in Ref.~\onlinecite{ponomarenko2011} both depend crucially on the screening properties of the double-SLG system, in particular, the suppression of the impurity-induced puddles in the studied layer due to the screening induced by the control layer, as noted already in Ref.~\onlinecite{kechedzi2012} using a perturbative analytical approach of double-SLG screening.  Since our current work is precisely on the non-perturbative screening properties of double-layer graphene system, we are in a good position to shed light on the experimental situation studied in Ref.~\onlinecite{ponomarenko2011}.
\begin{figure}[!htb]
 \begin{center}
  \centering
   \includegraphics[width=8.5cm]{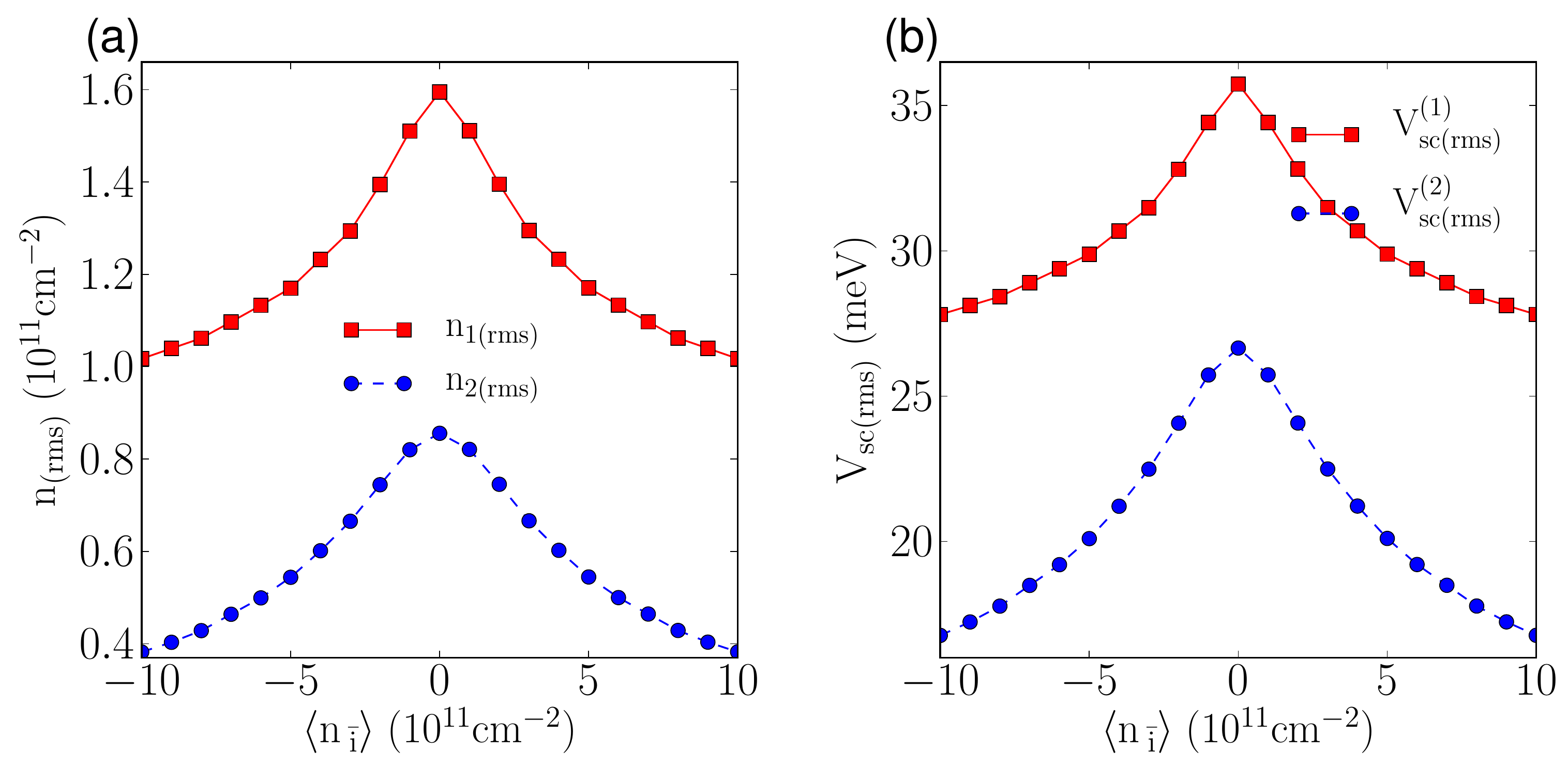}
   \caption{
           (Color online). Plots of (a) $\nrms$ and, (b) $\vrms$ at the CNP in layer ``i'' as a function of the doping in the other layer $\langle \rm{n}_{\bar {\rm i}}  \rangle$, for d$ = 1$nm, d$_{12} = 1$nm, and $\nimp= 3 \times 10^{11}$ cm$^{-2}$, for the gapless SLG-SLG heterostructure. The squares correspond to the bottom SLG layer and the circles correspond to the top SLG layer.
        } 
  \label{fig:figure_32}
 \end{center}
\end{figure}
Our results show that the two graphene layers forming the SLG-SLG heterostructure have in general
very different disordered ground states. 
This is exemplified by Figs.~\ref{fig:figure_32} and \ref{fig:figure_32g}.
Fig.~\ref{fig:figure_32} shows $\nrms$ and $\vrms$ at the CNP in layer ``i'' as 
a function of the doping in the other layer, layer $\bar {\rm i}$. We see that 
the effect of the doping in the control layer is very different if the 
studied layer is the top (2) or the bottom (1). In other words, the screening properties of the double-SLG heterostructure are highly asymmetric, as already noted in Ref.~\onlinecite{kechedzi2012} using a simple analysis, with the screening of the bottom layer by the top layer being very different quantitatively from the screening of the top layer by the bottom layer. 
This is due to the fact that the charge impurities are not distributed symmetrically, in particular we assumed that most of the charge impurities are closed to the surface of the \sio since hBN is much cleaner than \sio in terms of impurity disorder (see Fig.~\ref{fig:setup}). 
The main qualitative feature that we want to
emphasize is that the higher the disorder potential, $\vrms$, the higher
is $\nrms$ and therefore the lower is the resistivity, in contrast to normal
metals for which an increase of disorder corresponds to a resistivity increase.
The results of Fig.~\ref{fig:figure_32} support the scenario presented
in Ref.~\onlinecite{dassarma2012} provided our model for the gapless asymmetric double-SLG heterostructure applies to the experimental situation.

Fig.~\ref{fig:figure_32g} shows $\nrms$ and $\vrms$ in the bottom
(top) layer at the CNP as function of the doping in the top (bottom) layer 
for the case in which the graphene spectrum has a gap equal to 20~meV arising from the explicit presence of hBN substrate which might break the SLG sublattice symmetry as discussed in section II and as described by Eq.~\ref{eqn:gappedslg}.
Qualitatively the results are similar to the ones shown in Fig.~\ref{fig:figure_32}:
the layer with strongest disorder has the highest $\nrms$ and therefore
is expected to be more metallic than the cleaner layer.

\begin{figure}[!ht]
 \begin{center}
  \centering
   \includegraphics[width=8.5cm]{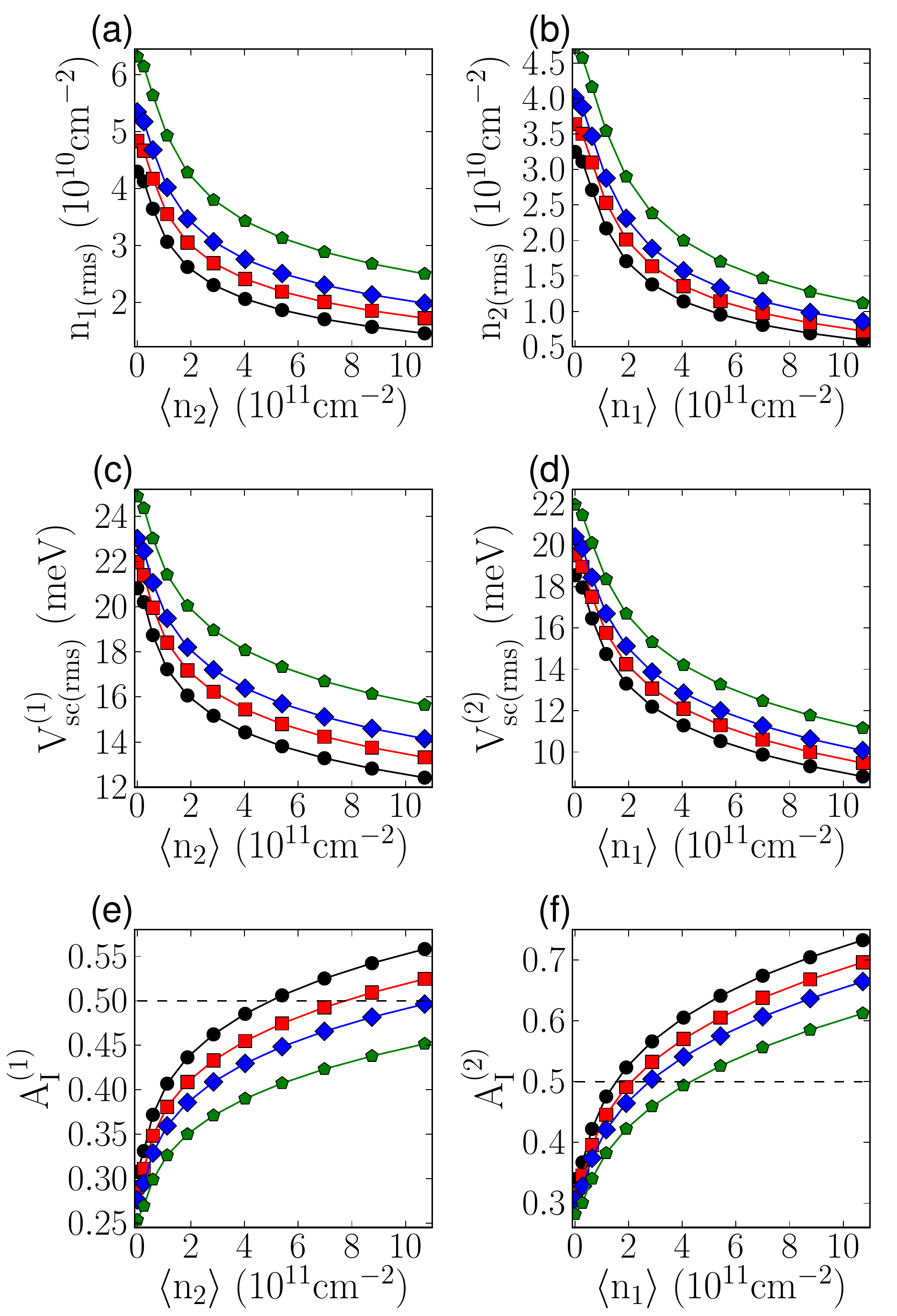}
   \caption{
           (Color online). Plots of (a) $\nrmsone$, (c) $\vrmsa$, and (e) A$_I^{(1)}$ as a function of $\navb$, at CN in the bottom layer,  and plots of (b) $\nrmstwo$, (d) $\vrmsb$, and (f) A$_I^{(2)}$ as a function of $\nava$ at CN in the top layer, all for d$= 5$ nm, d$_{12}= 1$ nm, and for different impurity strengths. The circles correspond to $\nimp = 1.5 \times 10^{11}$ cm$^{-2}$, the squares correspond to $\nimp = 1.75 \times 10^{11}$ cm$^{-2}$, the diamonds to $\nimp = 2 \times 10^{11}$ cm$^{-2}$, and the pentagons to $\nimp = 2.5 \times 10^{11}$ cm$^{-2}$,
        } 
  \label{fig:figure_34}
 \end{center}
\end{figure}

For SLG-SLG heterostructures for which the graphene spectrum has gap $\Delta$
it is interesting to consider impurity densities such that $\vrms\lesssim\Delta$.
In this situation we can have ground state configurations for 
which the majority of the studied layer is covered by insulating regions.
Under these conditions the layer
is expected to behave as a (bad) insulator \cite{rossi2011}.
It is therefore interesting to see how the fraction of the sample, $A_I$,
covered by insulating region in the studied layer at the CNP
depends on the doping in the control layer for impurity densities
such that $\vrms\sim\Delta$.
This is shown in Fig.~\ref{fig:figure_34}.
As the doping in the control layer increases the 
screened disorder in the studied layer decreases,
Figs.~\ref{fig:figure_34}~(c),~(d). As a consequence $\nrms$, i.e.
the amplitude of the carrier density inhomogeneities also decreases,
Figs.~\ref{fig:figure_34}~(a),~(b), so that in more regions
of the studied layer the effective local Fermi level falls within the band-gap.
We then see that, Figs.~\ref{fig:figure_34}~(e),~(f), as the doping
in the control layer increases, $A_I$ increases and, above a threshold,
reaches 50\%. For dopings in the control layer higher than this
threshold value there will not be a percolating path and the 
studied layer is expected to exhibit an insulating behavior.
The results of Fig.~\ref{fig:figure_34} therefore suggest a third
plausible scenario to explain the experimental results of
Ref.~\onlinecite{ponomarenko2011}: in the presence of a band-gap
in the graphene spectrum 
\cite{hunt2013,amet2013} the doping in the control layer,
by reducing the strength of the disorder in the studied layer, 
can drive it into a ground state in which more
than half of the area is insulating and therefore 
into an insulating state. This scenario
can be considered a generalization to the case when a finite
band-gap is present of the scenario presented in Ref.~\onlinecite{dassarma2012}.
In this scenario, where the interplay between the SLG band-gap
introduced by hBN and the disorder screening by the double-SLG
structure dominates transport properties in the system, there is a
density-tuned an effective metal-insulator transition from a gapped
insulator to an effective metal due to the percolation transition.
This is akin to the situation in gapped BLG \cite{rossi2011} 
where the opening of the single-particle gap has a different physical origin.

One important aspect of the results of Fig.~\ref{fig:figure_34} is that, as
in the experiment, for the layer with the lower effective disorder (higher mobility), in our case 
the top layer, the threshold value of the doping in the control layer
that drives it to be insulating is lower than for the more disordered 
layer (lower mobility).
The values of $\nimp$ and $d$ used to obtain the results
of Fig.~\ref{fig:figure_34} using the effective medium theory valid
for inhomogenous graphene ground states \cite{rossi2009} 
give values of the mobility that are of the same order, $10^5~{\rm cm^2/V\cdot s}$,
as observed in Ref.~\onlinecite{ponomarenko2011}.
It is therefore interesting to notice that for these values of $\nimp$
we find threshold values for the doping in the control layer that
are very close to the ones ($\sim 3\times 10^{11}{\rm cm^{-2}}$)
observed in Ref.~\onlinecite{ponomarenko2011}. Thus, it appears that the presence of an SLG gap coupled with the effective screening of the disorder in the studied layer by the tuning of the density in the control layer may very well be the physics dominating the observations in Ref.~\onlinecite{ponomarenko2011} although more experimental work will be necessary to clarify the situation.

The main difference between our results
and the results of Ref.~\onlinecite{ponomarenko2011}
is that in \onlinecite{ponomarenko2011} the top layer has a
higher effective disorder, lower mobility, than the bottom layer whereas 
our results show that the top layer always has a lower effective
disorder than the bottom layer, a consequence of the fact that
we assumed the charge impurities to be concentrated on the surface
of \sio, below the bottom layer. 
In our scenario for the MIT, this discrepancy would be resolved by assuming
that in the experiment of Ref.~\onlinecite{ponomarenko2011} the number
of charge impurities closer to the top layer is higher than 
in the bottom layer, perhaps due to the fabrication process or
to impurities adsorbed by the open  surface of the top layer.
Future experimental work with better control over the spatial location and magnitude of the impurity disorder should be able to resolve this issue completely and differentiate among the three distinct interpretations (i.e. Anderson localization, intrinsic thermal transport in clean graphene near the Dirac point, and a gap-induced metal-to-insulator transition as proposed in Refs.~\onlinecite{ponomarenko2011}, ~\onlinecite{dassarma2012}, and in the current work, respectively)  of the experimental observations in Ref.~\onlinecite{ponomarenko2011}.

%

\section{Discussion and conclusions}
\label{sec:conclusions}
%
In this work we have studied the effect of long-range disorder on the carrier distribution
density in graphene-based heterostructures. In particular, we have considered the 
case in which the main source of long-range disorder are charge impurities located
closed to the surface of the substrate. 
We have considered in detail three graphene-based heterostructures: 
(i)    SLG-SLG heterostructures formed by two sheets of single layer graphene separated by a dielectric film;
(ii)   BLG-SLG heterostructures formed by one sheet of bilayer graphene and one sheet of single layer graphene separated
       by a dielectric film;
(iii)  BLG-BLG heterostructures formed by two sheets of bilayer graphene separated by a dielectric film.

Our results show that, as for isolated graphenic layers, the presence of a long-range disorder potential
created by charge impurities induces long-range carrier density inhomogeneities, and in particular, these inhomogeneities break up the carrier density landscape into electron-hole puddles at the charge neutrality point. However, we find that the strength of these inhomogeneities, and of the screened
disorder potential, is in general much lower in the top layer due to the screening of the disorder
by the bottom layer, the one closer to the charge impurities. This is expected, but our results 
are the first to quantify such an effect for a large range of experimentally relevant conditions. 
In particular, our results show that in BLG-SLG heterostructures the strength of the screened
disorder in the SLG sheet is much lower than in the top SLG sheet of a SLG-SLG heterostructure.
This is due to the fact that at low energies, for most experimentally relevant conditions,
BLG has a higher density of states than SLG and therefore is much more efficient in screening
the top layer from the disorder. This also suggests that a very effective way to reduce the 
effect of charge impurities in SLG, or BLG, would be to reduce the thickness of the dielectric between
the graphenic layer and the metallic back gate.

One difficulty to obtain an accurate characterization, in the presence of charge impurities, of the carrier density profile 
of heterostructures comprising BLG is the fact that the impurities, and the carriers in the nearby graphenic layers and
metal gates, create an electric field with a component perpendicular to BLG that induces the opening of
band-gap ($\Delta$) in BLG. As a consequence, for heterostructures in which BLG is present,
the carrier density profiles and the BLG band-gap have to be calculated self-consistently.
Our results show that in general the average band gap $\Delta$ is not negligible.
For the set of parameters that we have used we find that the local value of $\Delta$
can be of the order of $50$~meV, the average $\Deltaav$ is of the order of 10-15~meV, 
and that for most of the cases the root mean square of $\Delta$, $\Deltarms$, is of the
order of $\Deltaav$, indicating that the inhomogeneities in the profile of $\Delta(\rr)$
are very strong. We expect these results to be very important to interpret transport
measurements in BLG-based heterostructures. 

We have also calculated the correlation ($C_{12}$) between the density profile in
the bottom layer and the one in the top layer. We find that for all the heterostructures
and conditions considered the two inhomogenous density profiles are correlated,
meaning that $C_{12}$ is positive and different from zero. This is due to the fact
that we assumed that the dominant source of long-range disorder are charge
impurities placed close to the bottom layer of the heterostructure.
Our results are important because provide a critical element for the interpration
of the recent results on the drag resistivity in SLG-SLG heterostructures
\cite{gorbachev2012,song2012,song2013}.

Our results are also directly relevant to the recently observed metal insulator
transition in graphene layers forming a SLG-SLG heterostructure.
In particular our results show that the transition from metallic to insulating
in the studied graphene layer of the SLG-SLG heterostructure, as a function
of the doping in the control layer, can be explained as a percolation-like
transition driven by the reduction of the amplitude and size
of the electron-hole puddles induced by the additional screening of
the impurity charges in the control layer of the disorder potential.

In particular, we show that the possible presence of an SLG gap, caused by the hBN substrate, could easily lead to the observed metal-insulator transition in the system as the charged disorder in the studied layer in suppressed due to screening induced by the control layer through density tuning.

The results presented are directly relevant to imaging experiments,
like scanning tunneling microscopy experiments, and for the interpretion
of transport measurements. In particular, the results for systems formed
by BLG, by providing both the strength of the band-gap induced by the 
perpendicular electric field generated self-consistently by the distribution
of charges in the heterostructure, and the strength of the screened
disorder potential, allow to identify the parameter regimes where
the BLG sheet is expected to behave as a bad metal or
as a bad insulator \cite{rossi2011}.
Our results are also important to better understand the
conditions necessary for the establishment of collective ground
states that have been theoretically predicted for 
SLG-SLG \cite{zhang2008jog,min2008},
BLG-SLG \cite{jzhang2013},
and BLG-BLG \cite{perali2013}
heterostructures.

\section{Acknowledgments}
\label{sec:Acknowledgments}
%
This work was supported by ONR, Grant No. ONR-N00014-13-1-0321, 
ACS-PRF Grant No. 53581-DNI5
and the Jeffress Memorial Trust. 
MRV acknowledges support from the Secretar\'ia de Educaci\'on P\'ublica, M\'exico. 
Computations  were  carried  out  on  the  SciClone Cluster at the College of William \& Mary.
%

\bibliographystyle{apsrev}

\end{document}